\documentclass[longauth]{aa} 
\usepackage[dvipsnames,svgnames]{xcolor}
\usepackage{graphicx}
\usepackage{subcaption}
\usepackage{lscape}
\usepackage{longtable}
\usepackage{multirow}
\usepackage{gensymb}
\usepackage{adjustbox}
\usepackage{amsmath}
\usepackage{ragged2e}
\usepackage{amssymb}
\usepackage{rotating}
\usepackage{xspace}
\usepackage{makecell}
\usepackage{supertabular}
\usepackage{multicol}
\bibpunct{(}{)}{;}{a}{}{,} 
\usepackage[colorlinks=true,allcolors=Blue]{hyperref}

\usepackage{txfonts}
\newcommand{\mycomment}[1]{}

\begin{document}

   \title{The SPHERE infrared survey for exoplanets (SHINE) \thanks{Based on data obtained with the ESO/VLT SPHERE instrument under programs SHINE (095.C-0298, 096.C-0241, 097.C-0865, 198.C-0209, 1100.C-0481, 1104.C-0416)  and snapSHINE (110.240D.001, 111.24QL.001).} 
   }
   \subtitle{V. Complete observations, data reduction and analysis, detection performances, and
    final results}

   \author{
%
A.~Chomez\inst{\ref{lesia}, \ref{ipag}} \and
P.~Delorme\inst{\ref{ipag}} \and
A.-M.~Lagrange\inst{\ref{lesia},\ref{ipag}} \and
R.~Gratton \inst{\ref{padova}} \and
O.~Flasseur\inst{\ref{cral}} \and
G.~Chauvin\inst{\ref{oca}} \and
M.~Langlois\inst{\ref{cral}} \and
J.~Mazoyer\inst{\ref{lesia}} \and
A.~Zurlo\inst{\ref{diegoportales1},\ref{diegoportales2}} \and
S.~Desidera\inst{\ref{padova}} \and
D.~Mesa\inst{\ref{padova}} \and
%
%
%
M.~Bonnefoy\inst{\ref{ipag}} \and
M.~Feldt\inst{\ref{mpia}} \and 
J.~Hagelberg\inst{\ref{geneva}} \and
M.~Meyer\inst{\ref{umich},\ref{eth}} \and 
A.~Vigan\inst{\ref{lam}} \and
C. Ginski\inst{\ref{galway}} \and 
M.~Kenworthy\inst{\ref{leiden}} \and 
%
%
%
D.~Albert \inst{\ref{osug}} \and
S.~Bergeon \inst{\ref{ipag}} \and
J.-L.~Beuzit\inst{\ref{lam},\ref{ipag}} \and
B.~Biller\inst{\ref{ifa_uoe},\ref{mpia}} \and 
T.~Bhowmik\inst{\ref{diegoportales1},\ref{diegoportales2}} \and
A.~Boccaletti\inst{\ref{lesia}} \and
M.~Bonavita\inst{\ref{padova}} \and 
W.~Brandner\inst{\ref{mpia}} \and 
F.~Cantalloube\inst{\ref{ipag}} \and 
A.~Cheetham\inst{\ref{geneva}} \and
V.~D'Orazi\inst{\ref{padova}} \and
C.~Dominik\inst{\ref{ams}} \and
C. Fontanive\inst{\ref{ifa_uoe},\ref{padova}} \and 
R.~Galicher\inst{\ref{lesia}} \and 
Th.~Henning\inst{\ref{mpia}} \and
M.~Janson\inst{\ref{mpia},\ref{stock}} \and
Q.~Kral\inst{\ref{lesia}} \and 
E.~Lagadec\inst{\ref{oca}} \and
C.~Lazzoni\inst{\ref{padova}} \and
H.~Le Coroller\inst{\ref{lam}} \and 
R.~Ligi\inst{\ref{oca}} \and 
A.-L.~Maire\inst{\ref{ipag}} \and 
G.-D.~Marleau\inst{\ref{mpia},\ref{due}} \and 
F.~Menard\inst{\ref{ipag}} \and
S.~Messina\inst{\ref{catania}} \and
N.~Meunier\inst{\ref{ipag}} \and  
C.~Mordasini\inst{\ref{bern}} \and 
C.~Moutou\inst{\ref{irap},\ref{lam}} \and
A.~M\"uller\inst{\ref{mpia}} \and
C.~Perrot\inst{\ref{lesia}} \and
M.~Samland\inst{\ref{mpia},\ref{stock}} \and
H.~M.~Schmid\inst{\ref{eth}} \and
T.~Schmidt\inst{\ref{lesia}} \and
V.~Squicciarini\inst{\ref{lesia},\ref{padova}}\and
E.~Sissa\inst{\ref{padova}} \and
M.~Turatto\inst{\ref{padova}} \and
S.~Udry\inst{\ref{geneva}} \and
%
L.~Abe\inst{\ref{oca}} \and
J.~Antichi\inst{\ref{padova}} \and
R.~Asensio-Torres\inst{\ref{mpia}} \and
A.~Baruffolo\inst{\ref{padova}} \and
P.~Baudoz\inst{\ref{lesia}} \and
J.~Baudrand\inst{\ref{lesia}} \and
A.~Bazzon\inst{\ref{eth}} \and
P.~Blanchard\inst{\ref{lam}} \and
A.~J.~Bohn\inst{\ref{leiden}} \and
S.~Brown~Sevilla\inst{\ref{mpia}} \and
M.~Carbillet\inst{\ref{oca}} \and
M.~Carle\inst{\ref{lam}} \and
E.~Cascone\inst{\ref{padova}} \and
J.~Charton\inst{\ref{ipag}} \and
R.~Claudi\inst{\ref{padova}} \and
A.~Costille\inst{\ref{lam}} \and
V.~De Caprio\inst{\ref{capodimonte}} \and
A.~Delboulb\'e\inst{\ref{ipag}} \and
K.~Dohlen\inst{\ref{lam}} \and
N.~Engler\inst{\ref{eth}} \and
D.~Fantinel\inst{\ref{padova}} \and
P.~Feautrier\inst{\ref{ipag}} \and
T.~Fusco\inst{\ref{onera},\ref{lam}} \and
P.~Gigan\inst{\ref{lesia}} \and
J.~H.~Girard\inst{\ref{stsci},\ref{ipag}} \and
E.~Giro\inst{\ref{padova}} \and
D.~Gisler\inst{\ref{eth}} \and
L.~Gluck\inst{\ref{ipag}} \and
C.~Gry\inst{\ref{lam}} \and
N.~Hubin\inst{\ref{eso_garching}} \and
E.~Hugot\inst{\ref{lam}} \and
M.~Jaquet\inst{\ref{lam}} \and
M.~Kasper\inst{\ref{eso_garching},\ref{ipag}} \and
D.~Le Mignant\inst{\ref{lam}} \and
M.~Llored\inst{\ref{lam}} \and
F.~Madec\inst{\ref{lam}} \and
Y.~Magnard\inst{\ref{ipag}} \and
P.~Martinez\inst{\ref{oca}} \and
D.~Maurel\inst{\ref{ipag}} \and
O.~M\"oller-Nilsson\inst{\ref{mpia}} \and
D.~Mouillet\inst{\ref{ipag}} \and
T.~Moulin\inst{\ref{ipag}} \and
A.~Origné\inst{\ref{lam}} \and
A.~Pavlov\inst{\ref{mpia}} \and
D.~Perret\inst{\ref{lesia}} \and
C.~Petit\inst{\ref{onera}} \and
J.~Pragt\inst{\ref{ipag}} \and
P.~Puget\inst{\ref{ipag}} \and
P.~Rabou\inst{\ref{ipag}} \and
J.~Ramos\inst{\ref{ipag}} \and
E.~L.~Rickman\inst{\ref{geneva}} \and
F.~Rigal\inst{\ref{ipag}} \and
S.~Rochat\inst{\ref{ipag}} \and
R.~Roelfsema\inst{\ref{nova}} \and
G.~Rousset\inst{\ref{lesia}} \and
A.~Roux\inst{\ref{ipag}} \and
B.~Salasnich\inst{\ref{padova}} \and
J.-F.~Sauvage\inst{\ref{onera},\ref{lam}} \and
A.~Sevin\inst{\ref{lesia}} \and
C.~Soenke\inst{\ref{eso_garching}} \and
E.~Stadler\inst{\ref{ipag}} \and
M.~Suarez\inst{\ref{eso_garching}} \and
Z.~Wahhaj\inst{\ref{eso_chile},\ref{lam}} \and
L.~Weber\inst{\ref{geneva}} \and
F.~Wildi\inst{\ref{geneva}}
}

\institute{
    LESIA, Observatoire de Paris, Universit{\'e} PSL, CNRS, Universit{\'e} Paris Cit{\'e}, Sorbonne Universit{\'e}, 5 place Jules Janssen, 92195 Meudon, France;  \label{lesia}\\
    \email{antoine.chomez@obspm.fr}
    \and
    Univ. Grenoble Alpes, CNRS-INSU, Institut de Planetologie et d'Astrophysique de Grenoble (IPAG) UMR 5274, Grenoble, F-38041, France; \label{ipag}
    \and
    INAF - Osservatorio Astronomico di Padova, Vicolo della Osservatorio 5, 35122, Padova, Italy \label{padova}
    \and
    CRAL, CNRS, Université Lyon 1,Université de Lyon, ENS, 9 avenue Charles Andre, 69561 Saint Genis Laval, France \label{cral}
    \and
    Universit{\'e} Côte d’Azur, OCA, CNRS, Lagrange, 96 Bd de l'Observatoire, 06300 Nice, France \label{oca}
    \and
    Aix Marseille Univ, CNRS, CNES, LAM, Marseille, France \label{lam} 
    \and
    $^{}$Instituto de Estudios Astrof\'isicos, Facultad de Ingenier\'ia y Ciencias, Universidad Diego Portales, Av. Ej\'ercito Libertador 441, Santiago, Chile \label{diegoportales1}
    \and
    $^{}$Millennium Nucleus on Young Exoplanets and their Moons (YEMS) \label{diegoportales2}
    \and
    Center for Space and Habitability, University of Bern, 3012 Bern, Switzerland \label{bern}
    \and
    Department of Astronomy, University of Michigan, Ann Arbor, MI 48109, USA \label{umich}
    \and
    Institute for Particle Physics and Astrophysics, ETH Zurich, Wolfgang-Pauli-Strasse 27, 8093 Zurich, Switzerland \label{eth}
    \and 
    Institute for Astronomy, University of Edinburgh, EH9 3HJ, Edinburgh, UK \label{ifa_uoe}
    \and
    Scottish Universities Physics Alliance (SUPA), Institute for Astronomy, University of Edinburgh, Blackford Hill, Edinburgh EH9 3HJ, UK \label{supa_edin}
    \and
    Centre for Exoplanet Science, SUPA, School of Physics \& Astronomy, University of St Andrews, St Andrews KY16 9SS, UK \label{supa_stand}
    \and
    Max Planck Institute for Astronomy, K\"onigstuhl 17, D-69117 Heidelberg, Germany \label{mpia}
    \and   
    Fakult\"at f\"ur Physik,
    Universit\"at Duisburg-Essen,
    Lotharstra\ss{}e 1,
    47057 Duisburg, Germany
    \label{due}
    \and
    Lunar and Planetary Laboratory, University of Arizona 1629 E. University Blvd. Tucson, AZ 85721, USA \label{tucson}
    \and
    Universit\"at T\"ubingen, Auf der Morgenstelle 10, D-72076 T\"ubingen, Germany \label{tueb}
    \and
    Unidad Mixta Internacional Franco-Chilena de Astronom\'{i}a, CNRS/INSU UMI 3386 and Departamento de Astronom\'{i}a, Universidad de Chile, Casilla 36-D, Santiago, Chile \label{umi}
    \and
    STAR Institute, University of Li\`ege, All\'ee du Six Ao\^ut 19c, B-4000 Li\`ege, Belgium \label{star}
    \and
    Centre for Exoplanet Science, University of Edinburgh, Edinburgh EH9 3FD, UK \label{ces_uoe}
    \and
    Department of Astronomy, Stockholm University, SE-10691 Stockholm, Sweden \label{stock}
    \and
    INAF - Catania Astrophysical Observatory, via S. Sofia 78, I-95123 Catania, Italy \label{catania}
    \and
    Univ. de Toulouse, CNRS, IRAP, 14 avenue Belin, F-31400 Toulouse, France \label{irap}
    \and
    ONERA (Office National dEtudes et de Recherches Arospatiales), B.P.72, F-92322 Chatillon, France \label{onera}
    \and
    European Southern Observatory (ESO), Karl-Schwarzschild-Str. 2,85748 Garching, German \label{eso_garching}
    \and
    Geneva Observatory, University of Geneva, Chemin des Mailettes 51, 1290 Versoix, Switzerland \label{geneva}
    \and
    Anton Pannekoek Institute for Astronomy, Science Park 9, NL-1098 XH Amsterdam, The Netherlands \label{ams}
    \and
    Leiden Observatory, Leiden University, PO Box 9513, 2300 RA Leiden, The Netherlands \label{leiden}
    \and 
    INAF - Osservatorio Astronomico di Brera, Via E. Bianchi 46, 23807 Merate, Italy \label{brera}
    \and
    INAF - Osservatorio Astronomico di Capodimonte, Salita Moiariello 16, 80131 Napoli, Italy \label{capodimonte}
    \and
    European Southern Observatory, Alonso de C\`ordova 3107, Vitacura, Casilla 19001, Santiago, Chile \label{eso_chile}
    \and 
    Space Telescope Science Institute, 3700 San Martin Drive, Baltimore, MD, 21218, USA \label{stsci}
    Instituto de F\'isica y Astronom\'ia, Facultad de Ciencias, Universidad de Valpara\'iso, Av. Gran Breta\~na 1111, Valpara\'iso, Chile \label{valpo1}
    \and
    N\'ucleo Milenio Formaci\'on Planetaria - NPF, Universidad de Valpara\'iso, Av. Gran Breta\~na 1111, Valpara\'iso, Chile  \label{valpo2}
    \and
    NOVA/UVA \label{nova}
    \and
    Université Grenoble Alpes, CNRS, Observatoire des Sciences de l’Univers de Grenoble (OSUG), Grenoble, France \label{osug}
    \and
    School of Natural Sciences, Center for Astronomy, University of Galway, Galway, H91 CF50, Ireland \label{galway}
}

\date{Received; accepted }
   \abstract
   {
   During the past decade, state-of-the-art planet-finder instruments like SPHERE@VLT, coupling coronagraphic devices and extreme adaptive optics systems, unveiled, thanks to large surveys, around 20 planetary mass companions at semi-major axis greater than 10 astronomical units. Direct imaging being the only detection technique to be able to probe this outer region of planetary systems, the SPHERE infrared survey for exoplanets (SHINE) was designed and conducted from 2015 to 2021 to study the demographics of such young gas giant planets around 400 young nearby solar-type stars. The analysis of the first part of the survey focused on 150 stars (SHINE F150) has already been published in a series of papers in 2021. An additional filler campaign called snapSHINE was conducted to acquire second epoch data using shallow observations.
   } 
   {
   In this paper, we present the observing strategy, the data quality, and the point sources analysis of the full SHINE statistical sample as well as snapSHINE. 
   }
   {
    Both surveys used the SPHERE@VLT instrument with the IRDIS dual band imager in conjunction with the integral field spectrograph IFS and the angular differential imaging observing technique. All SHINE data (650 datasets), corresponding to 400 stars, including the targets of the F150 survey, are processed in a uniform manner with an advanced post-processing algorithm called PACO ASDI. An emphasis is put on the classification and identification of the most promising candidate companions.
   }
   {
    Compared to the previous early analysis SHINE F150, the use of advanced post-processing techniques significantly improved by one or 2 magnitudes (x3-x6) the contrast detection limits, which will allow us to put even tighter constraints on the radial distribution of young gas giants. This increased sensitivity directly places SHINE as the largest and deepest direct imaging survey ever conducted. We detected and classified more than 3500 physical sources. One additional substellar companion has been confirmed during the second phase of the survey (HIP 74865 B), and several new promising candidate companions are awaiting second epoch confirmations.
   }
   {}  


\titlerunning{The final SPHERE infrared survey for exoplanets (SHINE). V.}

\maketitle
\section{Introduction and context}
\label{sec:intro}

The first exoplanet detection around a solar-type star \citep[51 Pegasi b,][]{Mayor_1995} was followed in the same year by the first detection of a substellar companion via direct imaging \citep[DI;][]{Nakajima_1995}, opening a new pathway towards exoplanet detection. Rapid instrumental progress and the advent of the first 8m class telescopes allowed imaging the first planetary-mass object around a brown dwarf \citep[2MASS~J1207334-393254~b;][]{Chauvin_2M1207} and a star \citep[AB~Pic~b;][]{Chauvin_abpic}, followed by the discoveries of HR~8799~bcde and $\beta$~Pictoris~b \citep{Marois_HR8799, Lagrange_2009}. These companions challenged the existing formation models, being much more massive than Jupiter ($\geq$ 5 $\text{M}_\text{Jup}$) and located much further away, at separations between 10 and several hundreds of astronomical units (au). Core accretion models \citep{Pollack_1996}, canonically used to explain the birth of our solar system, could not easily explain the formation of such objects, hinting towards other complementary formation mechanisms like gravitational instability inside the protoplanetary disk \citep{Cameron_1978}.

While radial velocity (RV) and transit techniques have been able to detect thousands of exoplanets at small separations, their ability to detect even giant planets is limited to semi-major axis less than 3 au with transit techniques, and less than 8-10 au \citep[see e.g.][]{Lagrange_rv_2023} with RV techniques, leaving high contrast imaging as the only method to reliably probe the outer region of planetary systems.
Microlensing surveys started to yield an increasing number of planetary-mass companions in the last decade \citep[see for instance Fig 1 of][]{Mroz_Poleski_2023_microlensing}, reaching $\sim$200 exoplanets mainly located in the 1-10 au range. This technique is uniquely sensitive to sub-Jovian exoplanets orbiting at separations of a few astronomical units, a population still out of reach of direct imaging. This method however relies on rare events, thus requiring decade-long surveys like the Optical Gravitational Lensing Experiment \citep[OGLE;][]{Udalski_OGLE} and is still not able to provide strong statistical results for massive Jovian planets \citep{Mroz_Poleski_2023_microlensing}. The future Roman space telescope will significantly increase the number of planets detected by microlensing, potentially up to a factor 10 \citep{Matthew_2019_Roman_microlensing, Samson_2020_Roman_microlensing}. As for astrometry, a small number of planetary candidates has been proposed using Gaia third data release  \citep[DR3,][]{GAIA_cat_planets}, though their parameters from this preliminary Gaia data may be affected by selection effects \citep[stars properties and/or orbital configurations, see e.g. the discussions in][]{Sozzetti_2023_gaia_planet}. Note that even the preliminary results already available from DR3 may be used to boost the yield from DI and improve our knowledge of systems found by other techniques \citep[see e.g.][]{Winn_2022,Philipot_2023b,Philipot_2023a,Winterhalder_GAIA_FRAVITY_2024}.   

DI faces two main challenges: the very small angular separations where companions are located with regard to their host star and the very high contrasts between planetary and stellar fluxes. Such limitations limit DI observations to close (typically less than 150 pc) young (less than 500 Myr) stars where the planets will be (angularly) separated from their parent star and warm enough to ensure a contrast detectable with current instruments.

Over the past two decades, high contrast imaging surveys looking for substellar companions were conducted \citep[for a more detailed overview, see][and references therein]{Chauvin_review_2018, Currie_PP7}. Starting with the first instruments with adaptive optics (AO) systems, typically targeting a few dozen stars \citep[see e.g.][]{Chauvin_MBM12_2002, Chauvin_Tuc_2003, Chauvin_VLT_CFHT_2006, Biller_survey_2007}, surveys progressively increased in size and sensitivity thanks to progress in instrumentation and image processing. Third-generation surveys, targeting $\sim$100 stars using coronagraphy and more advanced AO systems were able to probe the planetary regime, between 5 and 13 $\text{M}_{\text{Jup}}$, at a few dozens of au \citep[see e.g.][]{Janson_SEEDS, Desidera_NACOLP, Chauvin_NACO_LP, Vigan_NACO_LP, Galicher_IDPS}. The 4th and current generation surveys consider hundreds of stars: 400 for the SpHere INfrared survey for Exoplanets \citep[SHINE;][]{Chauvin_shine_2017}, 600 for the Gemini Planet Imager (GPI) Exoplanets Survey \citep[GPIES, the first half of which was published in][]{Nielsen_GPIES_2019}, and use dedicated planet-finder instruments equipped with extreme AO devices coupled with coronagraphy like the Spectro Polarimeter High-contrast Exoplanet REsearch \citep[SPHERE;][]{Beuzit_sphere} and GPI \citep{Macintosh_GPI}. Those surveys can efficiently probe the planetary regime beyond typically around 10 au \citep{Vigan_shine_2021, Nielsen_GPIES_2019}. Smaller surveys taking advantage of those facilities, more focused on very specific stellar populations, are also ongoing: for instance the B-star Exoplanet Abundance Study \citep[BEAST;][]{Janson_beast} in the Scorpius-Centaurus association (hereafter Sco-Cen). With the advent of Gaia \citep{gaia_dr2,gaia_dr3}, informed surveys are now conducted, compared to the past surveys that were blind, to try to maximize the detection efficiency of planetary companions \citep[typically in the order of $1\%$ for blind search, see e.g.][for a survey using this approach]{Currie_HCGA} with three recent planetary mass companion discoveries: AF~Lep~b \citep{Franson_af_lep, Mesa_af_lep, Derosa_af_lep}, HIP~99770~b \citep{Currie_HIP99770b} and Eps~Ind~b \citep{Matthews_2024_eps_ind_b}.

The SHINE survey was conducted over 200 nights between February 2015 and September 2021, using the GTO time allocated to the SPHERE consortium. This survey was designed to find and characterize new exoplanets or brown dwarfs (BD), so as to explore planetary architectures, determine the frequency of giant planets beyond 10 au, and study the dependency of this frequency with the spectral type of the host star. A last aim was to study the link between disks and planets to constrain planetary systems formation processes. 
The first series of papers presented a sample of the first 150 stars observed during the first half of the SHINE survey (hereafter called F150) were published in 2021 with the sample characteristics presented in \cite{Desidera_shine_2021} and the performances, data processing, and sources identification in \cite{Langlois_shine}. From these observations, a first statistical analysis has been conducted \citep{Vigan_shine_2021}, studying the frequency of giant planets, its dependency with regards to the stellar spectral types, and comparing the results to formation models. 
The first statistical results obtained using the F150 sample computed occurrence rates for planetary-mass companions around FGK stars of 0.7\% [0.3-2.9] 68\% confidence interval \citep[CI;][]{Vigan_shine_2021}, fully compatible with the GPIES survey and its upper limit at 6.7\% \citep[95\% CI interval;][]{Nielsen_GPIES_2019}. For higher mass stars (A and B type) SHINE and GPIES are also compatible with an occurrence rate of, respectively, 8.6\% [4.1–15.9] and 8.9\% [5.3–13.9] (68\% CI interval for both).
These results still suffer from large error bars and could be improved by using additional data. 

The present paper is part of a similar series with a first paper describing the sample (Paper IV, Desidera et al. in prep) and a third one studying the demographics and formation models (Paper VI, Chauvin et al. in prep). It adds 250 stars to the previously analyzed F150 sample to reach a total number of 400 stars, representing over 650 observations. This will be, alongside the GPIES survey \citep{Nielsen_GPIES_2019}, when fully released, the largest high contrast imaging exoplanet hunting survey ever conducted.
In addition to an increased number of targets, the present analysis uses an improved post-processing algorithm compared to the one used in \cite{Langlois_shine} \citep[SPECAL][]{Galicher_specal}: the PAtch-COvariance algorithm \citep[PACO;][]{Flasseur_paco, flasseur2020robustness, Flasseur_asdi}

The SHINE survey discovered two planets: HIP~65426~b \citep{Chauvin_HIP65426b} and PDS~70~b \citep[][although not part of the statistical sample presented here]{Keppler_PDS70_b_2018} and one brown dwarf \citep[HIP~64892~B;][]{Cheetham_BD} as well as, although the target selection was not optimized for it, 4 new disk detections \citep{Lagrange_disk_hd106906, Feldt_disk_hip73145,Sissa_2018_GSC07396,Perrot_disk_HD160305}.
While the SHINE survey focused on stars for which no stellar companions in the range of separation 0.1 to 6 arcsec at the epoch of its start (2014), a number of additional stellar companions in this separation range were discovered on SHINE data around 76 of the observed stars. Their observation is described in \cite{Bonavita_binaries}. These stars were not considered further in our analysis.

After its completion, 1220 SHINE companion candidates remained unconfirmed mainly identified in the IRDIS field-of-view (FoV) around 133 stars.
This large number of unconfirmed sources is due to an incomplete follow-up of detected sources due to losses of time because of a significant amount of poor observing conditions (SHINE was conducted in visitor mode) or scheduling constraints. Because both detections and non-detections are used in the statistical analysis \citep[see e.g.][]{Vigan_shine_2021}, their confirmation is critical to reducing biases and error bars.

A new and smaller survey called snapSHINE was proposed to the test the nature of 1010 candidates around 102 stars. The missing 31 stars harbor companion either too faint or too close to be accessible with snapSHINE. Finally, 76 out of the 102 targets were observed. The observing strategy of snapSHINE was designed to maximize the balance between observing time (as little as possible) and the number of sources to be re-observed (as much as possible). This snapshot technique has proven to be very effective to detect wide orbit exoplanets \citep{Bohn_2020_yses1b,Bohn_planets_2020,Bohn_2021_yses2b}. snapSHINE surveyed similar stars for which the candidates were observable using this approach. We should however acknowledge that this still leaves a number of potentially good but unluckily unconfirmed candidates, especially at short separations where snapSHINE has poor sensitivity and, moreover, where we might expect a larger prevalence of companions, as it will appear from Fig. \ref{fig:CMD_H23}-\ref{fig:CMD_K12}.

This paper is organized as follows: in Sect. \ref{sec:surveys} we present the SHINE and snapSHINE surveys with their associated observing conditions and in Sect. \ref{sec:data_reduction}, the data analysis and the algorithms used. The performances of the survey are presented in Sect. \ref{sec:sample_perf}, discussed in connection with regards to the observing conditions, and compared to the previous study. In Sect. \ref{sec:cc}, we present the detected point sources as well as the classification process used to identify them. Finally, in Sect. \ref{sec:ccl}, we present our conclusion and perspectives for future works.

\section{Surveys, related observing strategy and data quality}
\label{sec:surveys}
We present the observational strategy in Sect. \ref{subsec:obs_strat} for SHINE and in Sect.  \ref{subsec:obs_strat_snap} for snapSHINE. In Sect. \ref{subsec:overall_sample_quality}, we discuss the overall data quality of both surveys.

\begin{figure*}[t!]
    \includegraphics[width=\textwidth]{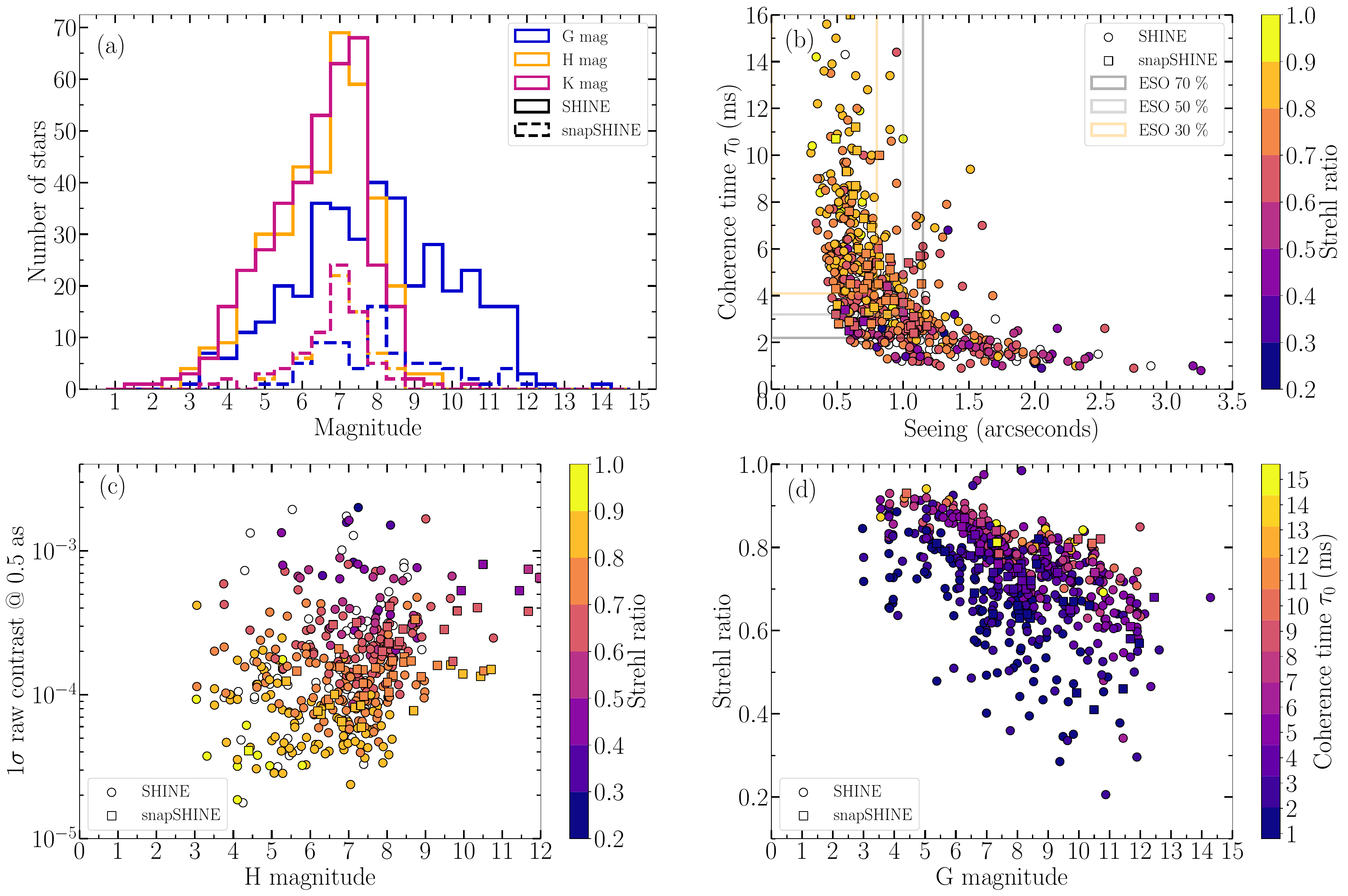}
    \caption{(a) Histogram of the star magnitudes (G, H, K) for the sample considered in this study. (b) Scatter plot showing the coherence time ($\tau_0$) against the seeing with the Strehl ratio color-coded. Data points without SPARTA Strehl information are colored in white. The grey box delimits the ESO $50\%$ best observing conditions (seeing < 1 as and $\tau_0 > 3.2$ ms) whereas the yellow box delimits the $30\%$ best (seeing < 0.8 as and $\tau_0 > 4.2$ ms). (c) Scatter plot showing the $1\sigma$ raw contrast at 0.5 arcsec (directly measured on the coronagraphic frame, before post-processing) with respect to the H-band (main IRDIS band used in SHINE) star apparent magnitude. The best contrast performances are achieved for the brightest stars with a high Strehl ratio. (d) Scatter plot displaying the Strehl ratio as a function of the star's apparent magnitude in the G band (AO spectral channel). As expected, the performances of the AO system deteriorate for fainter stars.}
    \label{fig:fig_obs_cond}
\end{figure*}

\subsection{SHINE observational strategy}
\label{subsec:obs_strat}

All SHINE observations were taken using standardized observational templates using SPHERE. Two modes were used, either  \verb+IRDIFS+ or \verb+IRDIFS-EXT+, both of which allow simultaneous observations by the Infra-Red Dual-beam Imager and Spectrograph \citep[IRDIS;][]{Dohlen_irdis} and the infrared Integral Field Spectrograph \citep[IFS;][]{Claudi_ifs, Mesa_ifs} in complementary wavelength bands. IRDIS was used in the dual-band imaging mode (DBI; \cite{Vigan_dbi}), either with the DB-H23 filter pairs in the case of the \verb+IRDIFS+ mode with IFS using the YJ bands, or with the DB-K12 filter pairs and YJH bands for IFS in the \verb+IRDIFS-EXT+ configuration. The latter configuration was mostly used for observations towards Sco-Cen (a 10-20 Myr association located at 100-150 pc) stars to maximize the detection capabilities of red young late L-type dusty planets \citep[e.g., HD 965086 b;][]{Chauvin_hd95086, Desgrange_HD95086}. 

\begin{figure}
    \centering
    \includegraphics[width = .9\linewidth]{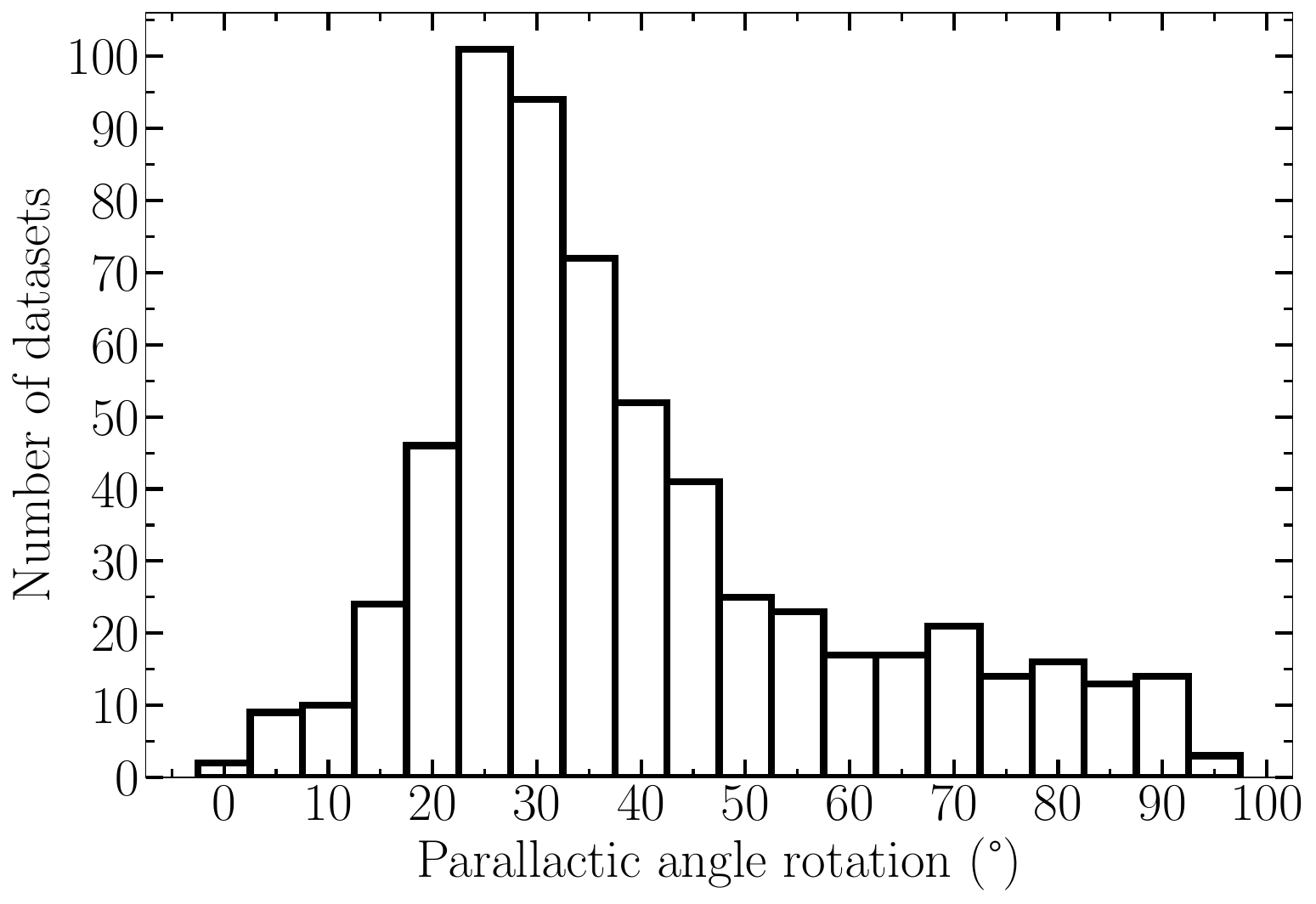}
    \caption{Histogram of the total amount of parallactic angle rotation during the coronagraphic sequences for the SHINE observations considered in this survey.}
    \label{fig:hist_rot}
\end{figure}

To efficiently schedule all of the SHINE observations and to maximize the amount of field of view rotation, critical for post-processing techniques relying on angular (and spectral) differential imaging \citep[A(S)DI;][]{Marois_adi}, the GTO team developed a dedicated optimal scheduler: \verb+SPOT+. This tool was described in \cite{Lagrange_spot} and in \cite{Langlois_shine}. It was used to optimize the observations for the whole survey while verifying specific constraints such as a variation of the parallactic angle of at least 20° and as close as possible to 30° as can be seen in Fig. \ref{fig:hist_rot}. Note that some datasets with small rotation angles are still used: for instance, two datasets of GJ 504 have around 5° of parallactic rotation but they still allow robust detection of the planet/brown dwarf (depending on the adopted age) GJ 504 b \citep[see e.g.][]{Kuzuhara_GJ504b_discovery, Bonnefoy_GJ504} because of its large angular separation (around 2.5 as) relative to its host star.

\begin{figure}
    \centering
    \includegraphics[width = .9\linewidth]{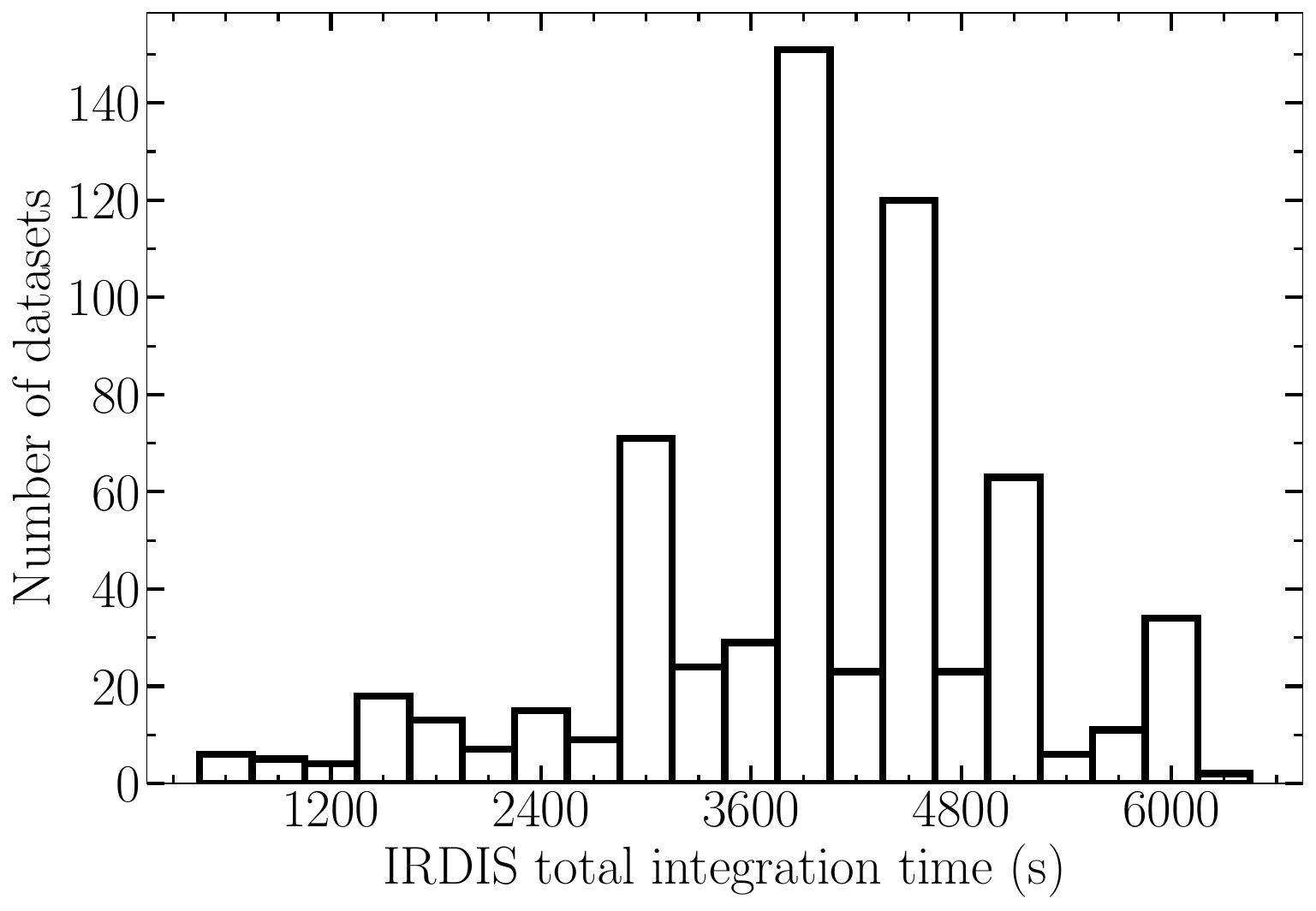}
    \caption{Histogram of the total coronagraphic integration time during the sequences for the SHINE observations considered in this survey.}
    \label{fig:hist_int_time}
\end{figure}

As previously mentioned, all SHINE observation sequences were standardized. They consist of an observation of the non-coronagraphic off-axis point-spread function (PSF) of the targeted star for flux calibration purposes, followed by a coronagraphic frame with four replicas of the star created by the deformable mirror \citep{Makidon_waffle} and used to check the centering of the star behind the coronagraphic mask (so-called waffle-frame produced by dedicated commands on the deformable mirror to create the four replicas). Once the star is centered, the science coronagraphic sequence is recorded. Finally, another waffle frame and another off-axis PSF are recorded, to estimate the astrometric and photometric changes between the beginning and the end of the observations, followed by a sky frame for calibration purposes. 
Some specific observations were taken with the waffle mode during the whole science sequence, mainly in the case of orbital monitoring of known companions. Indeed, such observations allow a frame-by-frame re-centering, thus greatly reducing the astrometric uncertainties due to any potential jitter of the telescope during the science acquisition. The standardization of most of SHINE observations also allows for a very consistent total integration time as can be seen in Fig. \ref{fig:hist_int_time}, with most of our observations achieved between 3600 s (the theoretical minimum observing time for a SHINE observation) and 5700 s of total science integration time (7200 s was the maximum time allowed). 

\subsection{snapSHINE observational strategy}
\label{subsec:obs_strat_snap}

This program was optimized to detect sources at a separation greater than 1 arcsec (corresponding to the 50-300 au range) with shallower observations compared to SHINE, with a typical sensitivity greater than 2 $\text{M}_{\text{jup}}$ at most. The observations are around 28 minutes long (including telescope overheads), with low field-of-view rotation, typically smaller than 4 degrees. Close or very faint sources (around 15\% of the total unconfirmed sources) were then out of reach of the survey. This trade-off in sensitivity was made to maximize the number of observations while still being sensitive to wide-orbit exoplanets such as those found by \cite{Bohn_planets_2020, Janson_b_cen_b, squicciarini21}.

The observations and calibration templates are similar to the SHINE survey with one minor change: the waffle mode is enabled for the whole coronagraphic sequence, to get the best astrometric precision and unambiguously identify background sources via the common proper motion test (see Sect. \ref{subsubsec:cpm_test}). As most unconfirmed candidates were detected using the \verb+IRDIFS+ setup, the latest is also used for the snapSHINE survey, with IRDIS using the BB\_H filter. The change from DB\_H23 to BB\_H filter for IRDIS was done to maximize the number of photons received (by increasing the bandwidth) given the relatively small exposure time of the snapSHINE observations.

In the present analysis, we consider all observations performed until the end of P111, even if the program has been extended to P113. A full analysis of the survey will be presented in a dedicated paper after its completion.

\mycomment{
\begin{figure}
    \centering
    \includegraphics[width = \linewidth]{figs/snapSHINE.pdf}
    \caption{Predicted average performance of snapSHINE compared to the SHINE nominal performances. All SHINE unconfirmed candidates (1220) are reported in light blue, the ones in green (1010) are the sources targeted by snapSHINE. The yellow dots are known exoplanets.}
    \label{fig:snapSHINE}
\end{figure}
}

\subsection{Overall data quality}
\label{subsec:overall_sample_quality}

The whole SHINE survey has been conducted in visitor mode, hence the quality of observations is highly variable. It is possible to retrieve information on these conditions either via the ESO MASS\footnote{\href{https://archive.eso.org/wdb/wdb/asm/mass_paranal/form}{https://archive.eso.org/wdb/wdb/asm/mass\_paranal/form}}-DIMM\footnote{\href{https://archive.eso.org/wdb/wdb/asm/dimm_paranal/form}{https://archive.eso.org/wdb/wdb/asm/dimm\_paranal/form}} or using the parameters related to image quality obtained by the real-time computer SPARTA \citep{Suarez_sparta} of the SAXO extreme adaptive optic of SPHERE \citep{Fusco_saxo, Beuzit_sphere}.
The seeing and the coherence time are retrieved using the ESO MASS-DIMM measurements.
We use the SPARTA Strehl ratio, not available for all observations: data are missing for the 115 observations taken between 2015-02-05 and 2015-05-19 and between 2016-06-02 and 2016-12-16. Note that the SPARTA Strehl can be slightly different from the ESO MASS-DIMM Strehl \citep[see for instance the Fig. 8 of][and discussion therein]{Milli_2017_SR}.
We can get information on the observing conditions through the prism of the ESO turbulence category in a similar fashion as in \cite{Courtney_barrer_HCI_model}, see Fig.  \ref{fig:fig_obs_cond}b. SHINE being conducted in visitor mode, we can assume that the observing conditions during the 200 nights were random. We should then expect that around 50\% of the observations were in the 50\% turbulence category (i.e. seeing < 1 as and $\tau_0 > 3.2$ ms). We found indeed that 48.35\% (311) of the observations are within this parameter space. 
66.14\% (424) of the SHINE observations are within the 70\% (seeing < 1.15 as and $\tau_0 > 2.2$ ms) turbulence category and 34.1\% (221) belong to the 30\% (seeing < 0.8 as and $\tau_0 > 4.2$ ms) category,
showing that the observing quality of a given SHINE observation behaves as expected from a collection of random observing nights. This impacts our ability to confirm faint candidate sources with a second epoch {if this was obtained in poorer conditions}, or on the contrary, we can achieve a deeper second epoch, unveiling new candidates undetected in the first epoch if the conditions were better.

Figure \ref{fig:fig_obs_cond}c-d also illustrates the degradation of the performances with increasing star magnitude. This is quantitatively shown in Fig. \ref{fig:fig_obs_cond}c which provides the $1\sigma$ raw contrast, computed on each coronagraphic frame, and then the median combined over the whole cube.

For snapSHINE, the survey was conducted in service mode, with a constraint of compliance with the 70\% turbulent category (seeing < 1.15 as and $\tau_0 > 2.2$ ms). About 98.7\% (75) of the observations were within the specified constraint. The observing conditions and data quality are summarized in Fig. \ref{fig:fig_obs_cond} alongside the SHINE's ones.

\section{Data reduction and analysis}
\label{sec:data_reduction}

\subsection{Calibrations}
\label{subsec:calib}

The standard calibration strategy described in \cite{Langlois_shine} was adopted throughout the entire survey. As for astrometry, we used the long-term calibration computed by \cite{Maire_calib} for all detected point sources.

\subsection{Pre-processing}
\label{subsec:preprocessing}

The pre-processing is identical to the one used for the F150, see \cite{Langlois_shine} for a detailed explanation of each step performed. In a nutshell, the first steps of the pre-reduction, consisting of dark, background, flat and bad pixels correction are performed at the High-Contrast Data Center \citep[HC-DC\footnote{\href{https://sphere.osug.fr/spip.php?rubrique16&lang=en}{https://sphere.osug.fr/spip.php?rubrique16\&lang=en}}, formerly known as SPHERE Data Center,][]{Delorme_sphereDC} by using the SPHERE data reduction and handling pipeline \citep[DRH;][]{Pavlov_DRH} provided by ESO. 

For IFS, a few additional steps are used to improve the wavelength calibration, implement a correction for cross-talk during the spectral extraction, and improve the bad pixels correction \citep{Mesa_ifs, Delorme_sphereDC}.

The only steps that differ from the F150 pre-reduction is the recentering of the individual frames. This was already presented in \cite{Chomez23_preparation, Dallant_Pacome}. In brief,  we fit a 2D Gaussian on each waffle pattern to estimate their subpixel location and infer the position of the star with subpixel precision, allowing thus precise recentering.
It has been shown in \cite{Dallant_Pacome} that this re-centering increases the signal-to-noise (S/N) of detected sources by a few percent compared to the previous pipeline. 

\subsection{ASDI post-processing}
\label{subsec:asdi_processing}

One critical part of processing such a survey is the homogeneity of the post-processing reduction and the ease of the analysis that comes afterward. During the first part of the SHINE F150 analysis \citep{Langlois_shine}, classical algorithms were used such as TLOCI \citep{Marois_TLOCI} and PCA \citep{Soummer_PCA, Mesa_ifs}, embedded into the \verb+SPECAL+ software \citep{Galicher_specal}. Those algorithms have the same three main drawbacks: (i) they provide non-statistically reliable detection limits (especially close to the star) thus leading in practice to many more false alarms than theoretically expected at the prescribed detection confidence, thus requiring (ii) a manual visual and time consuming source detection by an astronomer. Additionally (iii), differential imaging techniques heavily impact the shape of the off-axis PSF due to self-subtraction, introducing biases in the retrieve astro-photometry and requiring additional reduction steps (injection of fake planets for instance) to calibrate those biases.

To re-reduce all SHINE data (see Sect. \ref{subsec:nADI_processing} for the algorithm use for snapSHINE), we choose to use the PAtch COvariance \citep[PACO;][]{Flasseur_paco} algorithm as our baseline for both IFS and IRDIS in an automated pipeline detailed in \cite{Chomez23_preparation}. More specifically, we used \verb+robust PACO ASDI+ algorithm \citep{Flasseur_asdi, flasseur2020robustness}. The underlying models capture the statistics of the noise at a local scale of small patches through mixtures of scaled multi-variate Gaussians. The algorithm allows for a better estimation of the spatial and spectral correlation of the noise than classical methods. \cite{Chomez23_preparation} demonstrated on a sub-sample of the F150 that this approach improves the IRDIS contrast by one to two magnitudes (x3-x6), depending on the angular separation, compared to more classical ADI approaches like TLOCI and PCA. The gain for IFS is more modest because IFS data are more photon-limited than IRDIS. In both cases,  the detection limits provided by PACO are statistically reliable, thus allowing for a realistic false alarm rate.

The ASDI option of PACO is used to jointly process the temporal and spectral dimensions, producing spectrally combined S/N maps, and using spectral weights to maximize the detection of candidate sources having a similar spectral signature. These weights, with as many components as wavelength ($L=2$ for IRDIS and $L=39$ for IFS) are referred to as spectral priors. The selection, optimization process, and associated detection performances of the spectral priors are fully explained in \cite{Chomez23_preparation}.

\subsection{No-ADI post-processing}
\label{subsec:nADI_processing}

Because of the small number of frames and the small field-of-view rotation, we can not process the snapSHINE observations with PACO. We used instead the NoADI routine (a simple derotated median stack of all images) embedded in \verb+SPECAL+ \citep{Galicher_specal} for the IRDIS processing which is well adapted given the small $\Delta$PA variations, and as the sources of interest are located in the background limited region. Additionally, we processed the IFS PSFs using the algorithm described in \cite{Bonavita_binaries}. In a nutshell, this algorithm uses the off-axis PSFs taken before and after the coronagraphic sequence and computes a subtraction between the two after spectral-collapsing, normalization, and re-aligning of PSFs. In case of very bright sources (mainly binary stars) in the IRDIS FoV that can saturate the coronagraphic frames, and therefore cannot be characterized, we also use the no-ADI algorithm on the off-axis PSFs as they are often taken with a neutral density to avoid saturation. 4 binary stars are characterized using this method.

\section{Detection sensitivity and comparison with the  F150 analysis}
\label{sec:sample_perf}

In this section, we present the performances achieved for SHINE and snapSHINE. As a first step, we present the contrast curves for both surveys in Sect. \ref{subsec:perf_contrast} as well as a discussion on those limits. In the second part, in Sect. \ref{subsec:conversion_to_mass} we translate the contrast curves into detection limits expressed in Jupiter masses for the same stars considered in the F150 sample \citep[][]{Desidera_shine_2021, Vigan_shine_2021}, to quantify the improvement using \verb+PACO ASDI+ compared to the previous analysis \citep{Langlois_shine}.

\subsection{Contrast curves and performance comparison regarding observing conditions}
\label{subsec:perf_contrast}

\begin{figure*}[t!]
    \centering
    \includegraphics[width=\textwidth]{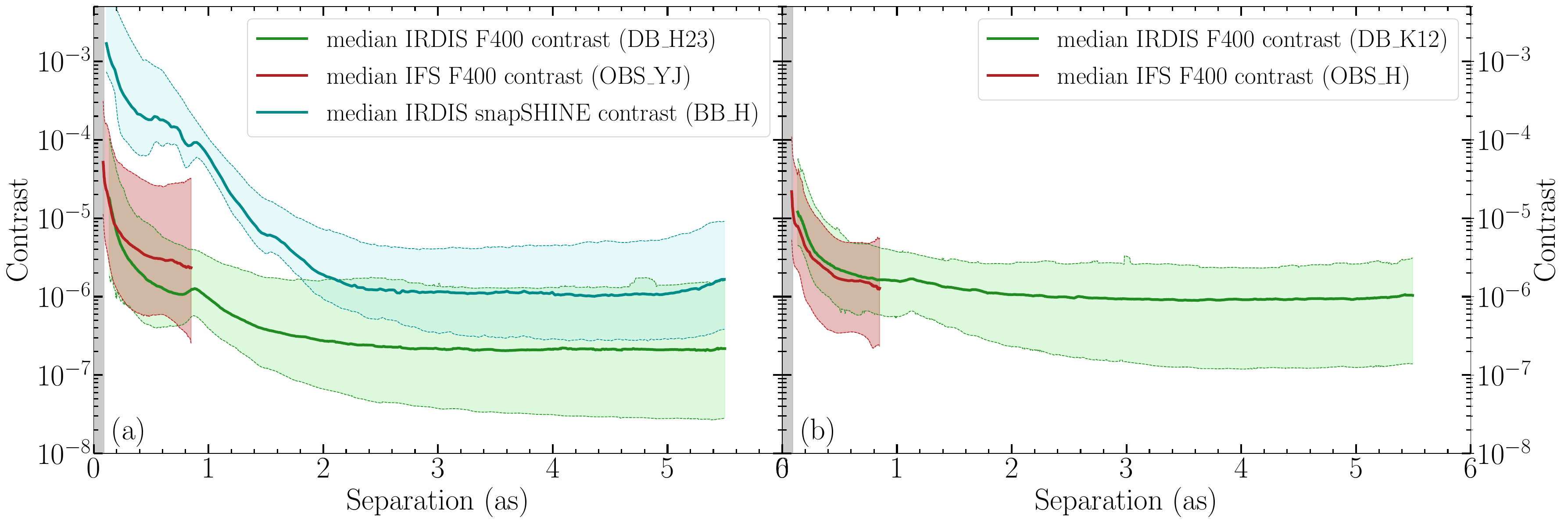}
    \caption{Contrast curves at $5\sigma$ obtained on the whole survey for both instruments, IRDIS in green and IFS in red. (a) IRDIFS setup. (b) IRDIFS-EXT setup. The contrast curves of the snapSHINE survey are represented in blue. The grey overlaid rectangles at small angular separations show the average inner working angle of the coronagraphic masks.} 
    \label{fig:contrast_curves}
\end{figure*}

The achieved performances in terms of contrasts are presented in Fig. \ref{fig:contrast_curves} for both the \verb+IRDIFS+ mode and the \verb+IRDIFS-EXT+ mode. 

Compared to the F150 analysis \citep{Langlois_shine}, the IRDIS performances with the \verb+IRDIFS+ setup are significantly improved, reaching a median contrast of $10^{-6}$ at 0.8 arcsec and $2\times10^{-7}$ beyond  2.5 as. The gain for the IFS data is not as high, only slightly improving the detection limits. However, the \verb+SPECAL+ detection curves are not reliable with regards to the false alarm probability at $5\sigma$, when \verb+PACO+ is, especially at short separation. Even if the contrast improvement is moderate, the statistical reliability of the contrast curves derived by \verb+PACO+, crucial for the statistical analysis, is much stronger. For the \verb+IRDIFS-EXT+ setup, the achieved contrast at short separations (below 1 as) is comparable to the one achieved in the \verb+IRDIFS+ setup. At larger separations (> 1 as), the contrast in the K band is limited by the strong thermal (background) noise impacting IRDIS images, thus reducing the achieved contrast compared to the H band, where this type of noise is significantly lower.

\begin{figure*}[t!]
    \centering
    \includegraphics[width=\textwidth]{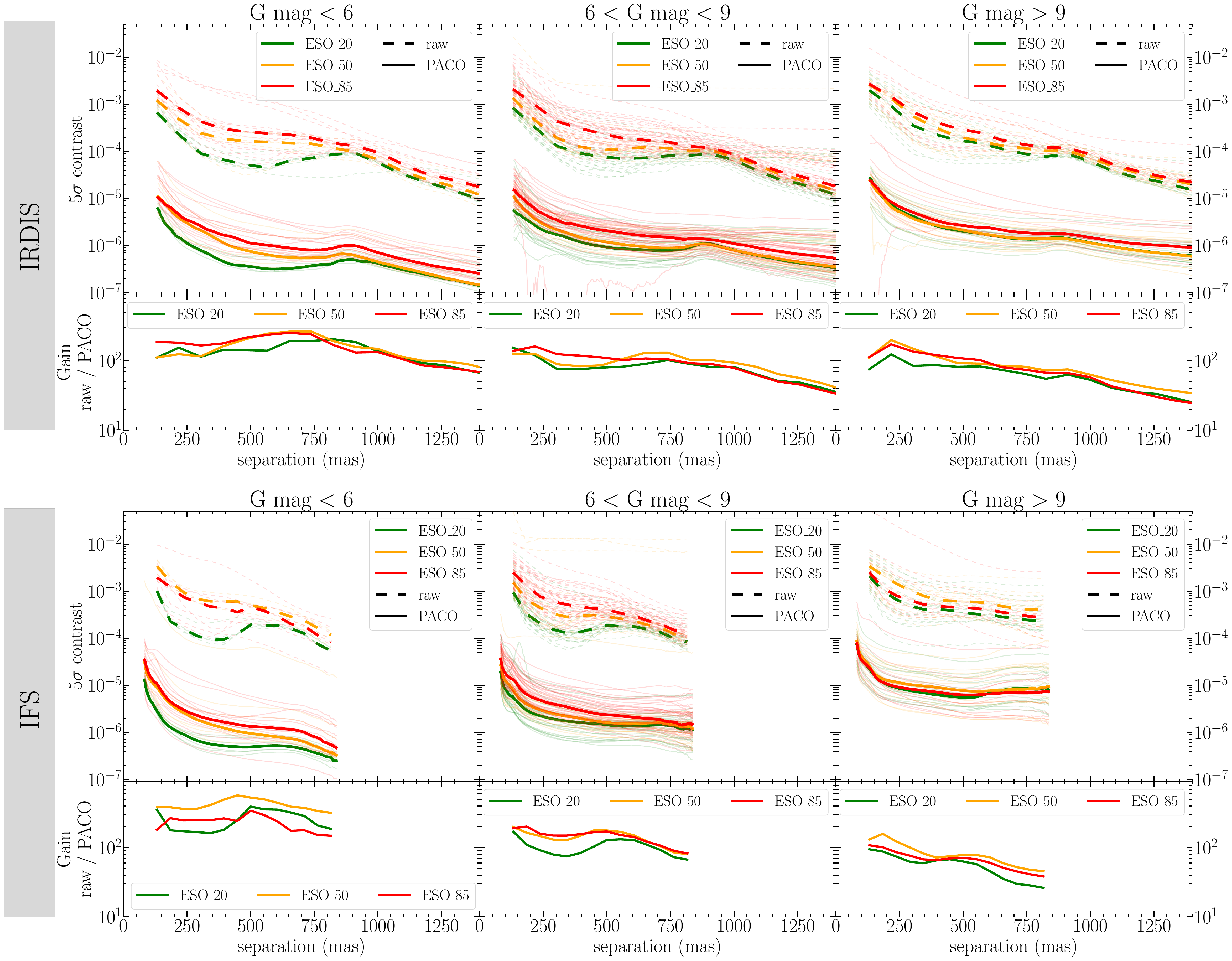}
    \caption{Contrast comparison between the raw contrast (median average spectrally and spatially, dashed lines) and the PACO contrast, both at $5\sigma$. 3 bins of star magnitudes and observing conditions are considered. Individual observations are represented by the thin lines and the median for each observing condition by the thick lines. The gain gives the contrast improvement between the median raw curves and the median PACO curves, provided for different observing conditions and different star brightness. The average size of the coronagraph's inner working angle is around 100 mas.} 
    \label{fig:contrast_comp}
\end{figure*}

Following \cite{Courtney_barrer_HCI_model}, we plot the raw and post-processing contrasts using bins of stellar G magnitude and the ESO SPHERE turbulence category classification. We consider  three categories: G $<$ 6, 6 $<$ G $<9$ and G $>$ 9. All datasets impacted by the low-wind effect \citep[windspeed $<3~\text{m.s}^{-1}$;][]{Milli_2018_LWE} have been removed from this study. Additionally, only datasets with a field-of-view rotation between 20° and 40°,  and a total integration time between 2700 and 5400s were considered. This parameter space is representative of the majority of our observations (see Figs. \ref{fig:hist_rot}-\ref{fig:hist_int_time}) while tight enough to mitigate the impact of those variables on the plots, allowing us to only consider the star brightness and the observing conditions\footnote{A more thorough study on a case by case basis could also be informative on the SPHERE performance itself but such study is out of the scope of this paper aiming at deriving general conclusions.}. Additionally, we estimate the gain provided by \verb+PACO ASDI+ by computing the ratio between the median raw curves and the median post-processed curves for each star brightness and observing category. As expected, the raw contrast performance (i.e., directly measured in the raw coronagraphic frames, in our case median-averaged spectrally and temporally over the whole 4D cube) decreases the fainter the star, as can be seen in Fig. \ref{fig:contrast_comp}. While on IRDIS, \verb+PACO+ improves the contrast by a factor around 100 for separations below 1 arcsec for all stars' brightnesses, the gain is much more dependant on the star brightness for IFS: for bright stars, the gain can go up a few hundreds, achieving better performances at small angular separations than IRDIS, but for faint stars, the gain is comparable or worse than for IRDIS. As shown in Fig. \ref{fig:contrast_comp}, for the bright and intermediate stars, we clearly see the improvement in contrast due to the AO correction between 200 and 500 mas (the dip in the raw contrast curves). In those cases, the data are still speckles-limited. However, this dip disappears for faint stars, becoming a plateau. This indicates a switch in the noise regime, from a contrast-limited one to a background-limited one. Given the already worse raw contrast in that case, it might explain why IFS can underperform compared to what is traditionally expected and explain the broader dispersion compared to IRDIS in the post-processed contrast seen in Fig. \ref{fig:contrast_curves}\footnote{This limitation in the results with the SPHERE IFS is due the fact that only the detector, and not the whole instrument is cooled. This choice was done by the builders to reduce costs and simplify the instrument, given that according to the original specifications, IFS was designed only for objects brighter than $J<8$}. 
We finally note the excellent performances of SPHERE coupled with \verb+PACO ASDI+ when observing bright stars under the best observing conditions; for example, contrasts better than $10^{-6}$ at 0.5 arcsec can be achieved.

\begin{figure*}[t!]
    \centering
    \includegraphics[width=.9\textwidth]{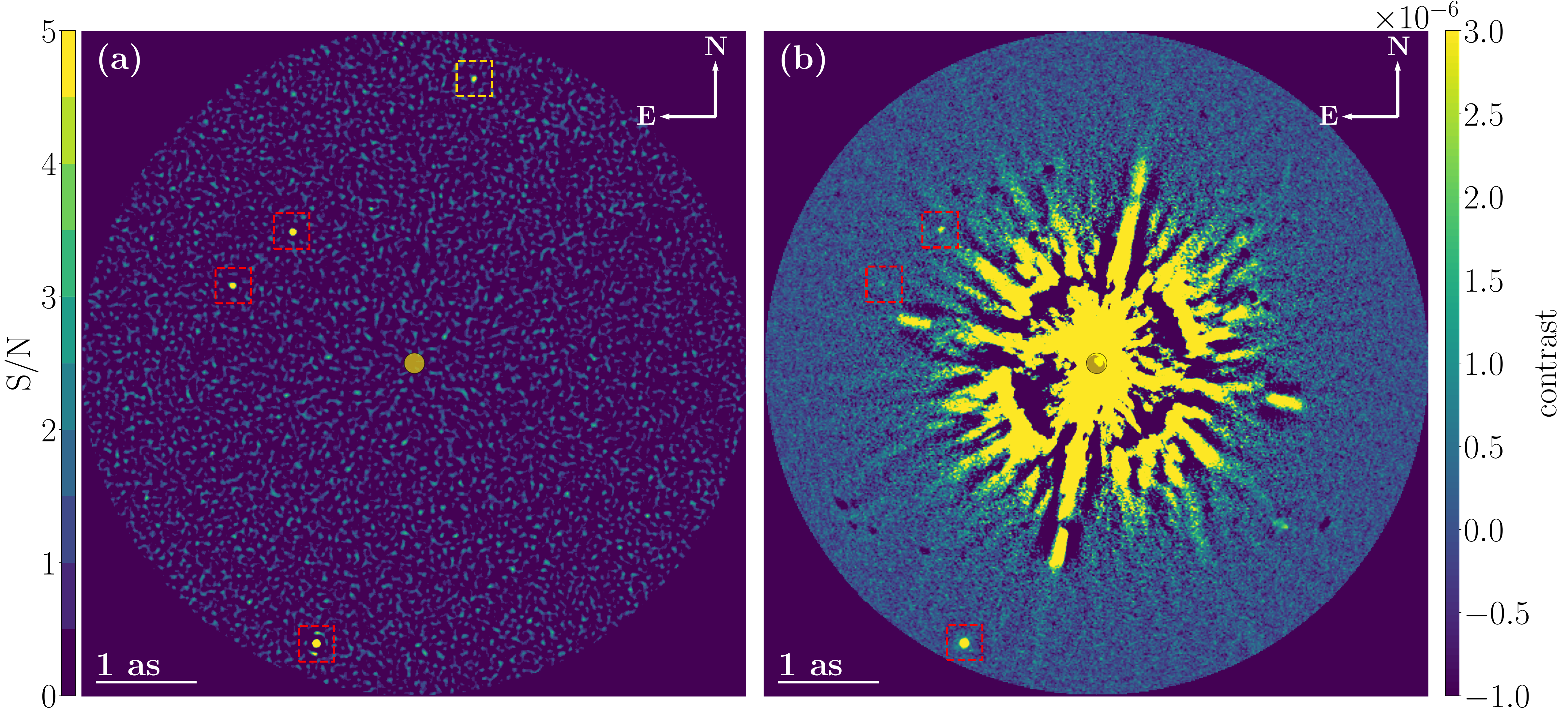}
    \caption{(a) S/N map of the star TYC\_8634\_1393\_1 with the SHINE observation using PACO ($\tau=5\sigma$) (b) Residual map of the same star with the snapSHINE observation using NoADI. 4 detections are present in the PACO S/N map, highlighted by the dashed squares. Only the 3 red ones are redetected in the snapSHINE observation (background sources). The other one in yellow is too faint to be detected by snapSHINE. A 15x15 pixels 2D median filter is applied on the snapSHINE residual map and the detection is performed manually.} 
    \label{fig:map_comp_shine_snapshine}
\end{figure*}

Finally, the contrast curves derived with \verb+SPECAL+ for snapSHINE are presented alongside the ones obtained with \verb+PACO ASDI+ in \verb+IRDIFS+ mode in Fig. \ref{fig:contrast_curves} for comparison. As expected, snapSHINE is much less sensitive below 2 arcsec but reaches the expected contrast in the background limited area, being only less sensitive than SHINE by about two magnitudes. Figure \ref{fig:map_comp_shine_snapshine} shows as an example a SHINE and a snapSHINE image of the same target. While the three brightest sources in the FoV are recovered in the snapSHINE image, the faintest one is not.

\subsection{Comparison with the F150 early analysis}
\label{subsec:conversion_to_mass}
In order to quantify the gain using PACO, we compared the detection limits obtained in this work to the one published in \cite{Vigan_shine_2021}, only using the stars included in the F150. 

\begin{figure*}
\centering
\begin{subfigure}{.49\textwidth}
  \centering
  \includegraphics[width=\linewidth]{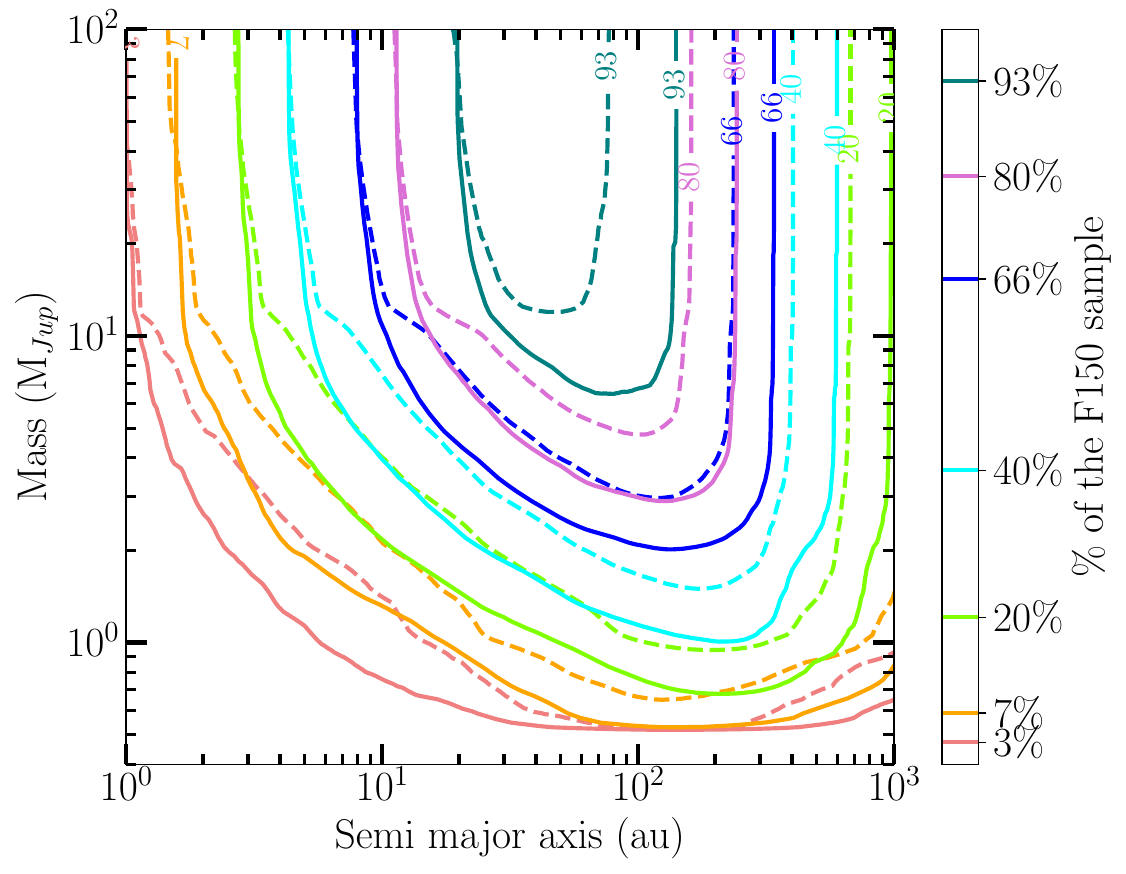}
\end{subfigure}%
\hfill
\begin{subfigure}{.49\textwidth}
  \centering
  \includegraphics[width=\linewidth]{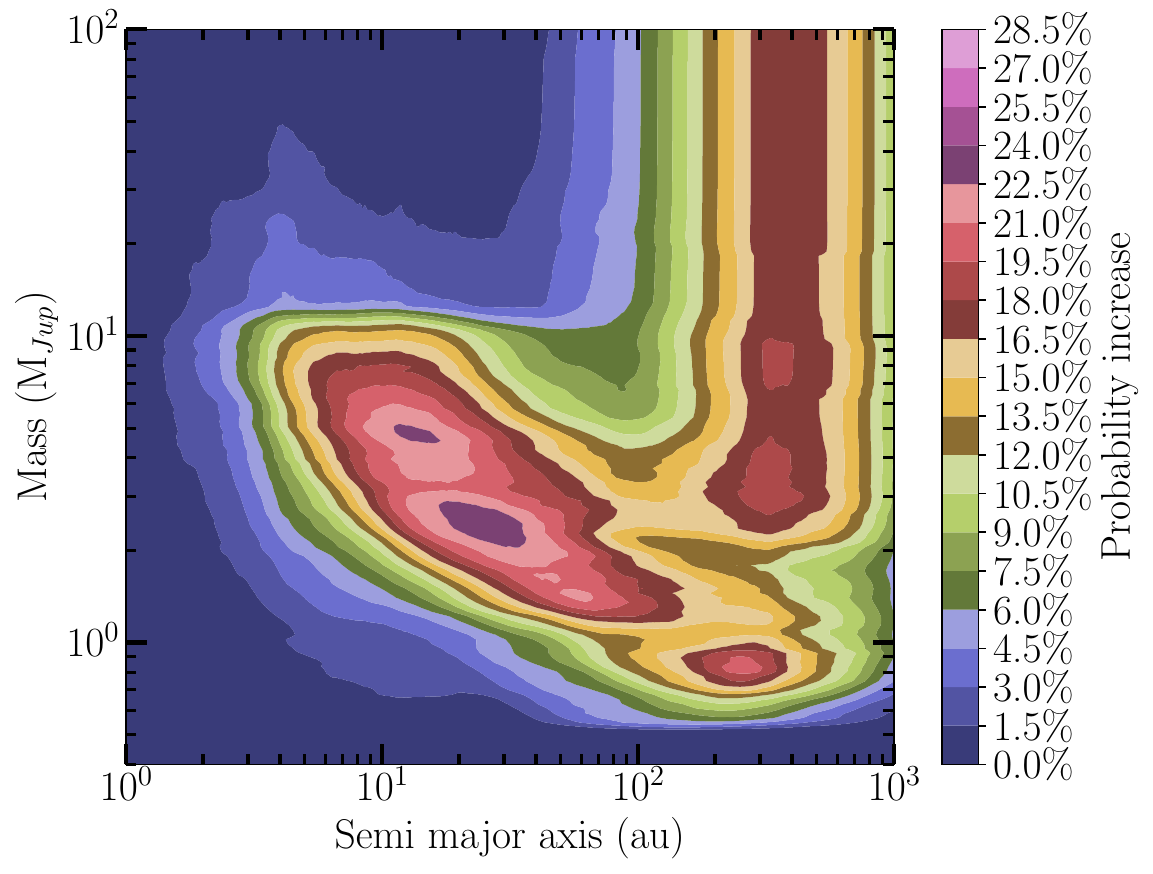}
\end{subfigure}
\caption{Left: Comparison between the sensitivity achieved by \cite{Vigan_shine_2021} using classical post-processing algorithms \citep{Langlois_shine} on the F150 (dashed lines) compared to the sensitivity achieved using PACO on the same sample of stars (plain lines). The lower limit seen around 100 au at 0.5 $\text{M}_{\text{Jup}}$ is due to the COND model, which does not provide inputs for lower masses. Right: Probability increase ($\mathcal{P}_{\text{PACO}}(\text{sma}, \text{M}_{\text{Jup}}) - \mathcal{P}_{\text{SPECAL}}(\text{sma}, \text{M}_{\text{Jup}})$) between the PACO analysis and the one presented in \cite{Vigan_shine_2021}.}
\label{fig:thumbs_up}
\end{figure*}

As in \cite{Vigan_shine_2021}, we used the 1D contrast curves and convert them into detection limits in Jupiter masses using a recent update of \verb+MADYS+\footnote{\href{https://github.com/vsquicciarini/madys}{https://github.com/vsquicciarini/madys}} \citep[v1.2,][]{madys}, with the ages extracted from \cite{Desidera_shine_2021} and the parallaxes and fluxes from SIMBAD. For the conversion, we used the COND2003 model \citep{Baraffe_COND2003}.
To convert the mass curves into detection limits, we used EXO-DMC\footnote{\href{https://github.com/mbonav/Exo_DMC}{https://github.com/mbonav/Exo\_DMC}} \citep{EXO_DMC}.
The grid parameters are designed to be similar to the ones used in \cite{Vigan_shine_2021} with the semi-major axis ranging from 0.1 to 1000 au (500 points, log uniform sampling) and the masses from 0.1 to 100 $\text{M}_{\text{Jup}}$ (200 points, log uniform sampling). A thousand draws are made for every cell in the grid with the inclination taken from a uniform distribution without any constraints and the eccentricity from a Gaussian law following $\mathcal{N}(\mu=0,\,\sigma=0.3)$. Figure \ref{fig:thumbs_up} clearly shows the gain in sensitivity obtained using \verb+PACO+ compared to the post-processing algorithms used in \cite{Langlois_shine}. The sensitivity is improved at all separations: for instance, the parameter range within which we are able to identify companions around 93\% of the sample (140 stars) region is greatly extended, reaching 6 $\text{M}_{\text{Jup}}$ at 100 au. At 10 au, while the previous analysis was able to reach 4 $\text{M}_{\text{Jup}}$ for 20\% (30 stars) of the sample when using \verb+PACO ASDI+, we achieve a detection limit of 2 $\text{M}_{\text{Jup}}$. We are even able to go beyond the 1 $\text{M}_{\text{Jup}}$ sensitivity for 20\% of the F150 sample at 70 au, performances that the \verb+SPECAL+ analysis could barely reach at 200 au. We then computed the subtraction between the \verb+PACO+ analysis and the analysis presented in \cite{Vigan_shine_2021}. This map is computed by subtracting, for every cell in the grid, the probability obtained using \verb+PACO+ $\mathcal{P}_{\text{PACO}}(\text{sma}, \text{M}_{\text{Jup}})$ by the one obtained using \verb+SPECAL+ $\mathcal{P}_{\text{SPECAL}}(\text{sma}, \text{M}_{\text{Jup}})$ and provide a clear view of the area of the parameters' space where we improve the most compared to the previous analysis.

\section{Candidate companion classification and analysis}
\label{sec:cc}

In this section, we summarize the confirmed planets or brown dwarf companions detected during the second part of the survey in Sect. \ref{subsec:confirm_sources}. Sub-stellar companions already presented in \cite{Langlois_shine} and stellar binaries studied in \cite{Bonavita_binaries} will not be discussed here but their astro-photometry extracted with PACO will still be included in the VizieR table linked to this article, as the whole survey has been re-processed. The source classification scheme used for the survey is presented in Sect. \ref{subsec:classification}. Online CDS tables with the astro-photometry of all IRDIS and IFS sources as well as all IFS spectrum for sub-stellar objects are linked to this article.

\subsection{Confirmed exoplanets and brown dwarfs}
\label{subsec:confirm_sources}

\begin{figure}[ht!]
    \centering
    \includegraphics[width = \linewidth]{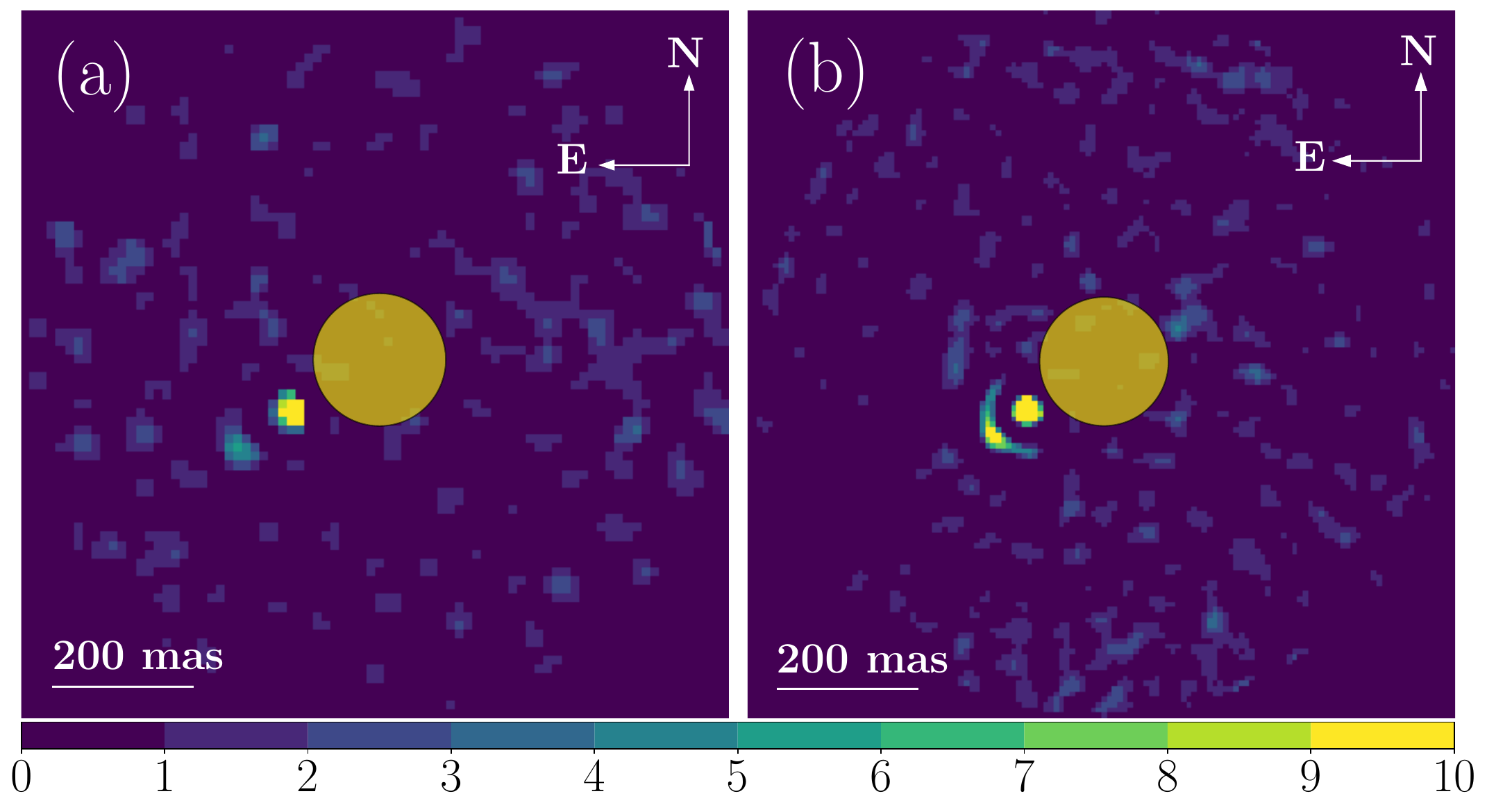}
    \caption{S/N maps for (a) IRDIS and (b) IFS of HIP 74865 B. The coronagraph is represented by the yellow circle. The brown dwarf is clearly visible southeast of it. Maps are cropped at a 500 mas radius. The color bar represents the S/N.}
    \label{fig:HIP_74865}
\end{figure}

One additional substellar companion has been confirmed during the second part of the SHINE survey: HIP 74865 B. Already detected in 2015 \citep{Hinkey_2015_BD} using Sparse Aperture Masking (SAM) on VLT/NACO \citep{Rousset_NAOS}, it is a brown dwarf of about 30-60 $\text{M}_{\text{Jup}}$ orbiting around a 15 Myr Sco-Cen star at about 20 au of its host star. The first image of the companion has been taken in the framework of the SHINE survey (see Fig. \ref{fig:HIP_74865}). Following this confirmation, two additional SPHERE epochs were recorded using the SAM mode of SPHERE \citep{Cheetham_SPHERE_SAM}. Those three observations and an in-depth analysis of this brown dwarf will be presented in a forthcoming paper (Cantalloube et al., in prep). The limited number of additional confirmed companions with respect to F150 should not be considered surprising because in the course of the SHINE survey higher priority was given to observations of the best targets - where detections were more probable.

\subsection{Classification of the detected point sources}
\label{subsec:classification}

In order to ensure the most reliable and complete statistical analysis that will come after this work, we need to classify all sources detected during the survey. This classification must be performed as thoroughly as possible by considering each and every source as a potential planetary candidate with the objective of leaving as few as possible candidates with an ambiguous status. Those ambiguous candidates are the ones impacting the statistical analysis as their nature (i.e., background contaminants or bound sources) cannot be accurately assessed.
We want to stress here that there are two ways of counting the detections: one where we sum all the detected point sources and one where we consider only the physical sources. For instance, there are six HR8799 datasets in the present analysis, where all four planets (on IRDIS) are detected on every epoch. The number of physical sources is four (one per planet) but the number of detection is 24. Combining over 700 observations, 7223 (5858 SHINE + 1365 snapSHINE)  point sources were detected. In the following, both numbers will be given if different. The lack of a second observation for every star as well as heterogeneous observing conditions calls for a complex identification scheme. Its various steps are described in the following subsections. As it requires a photometric and astrometric analysis of each detected point source as done for actual exoplanets, this process is particularly time-consuming. 

\subsubsection{Common proper motion test}
\label{subsubsec:cpm_test}

The most robust method to assess the physical link between a candidate companion and a star is the common proper motion test, because it does not make any hypothesis on its spectrum. It compares the motion of a detected point source between two observations and assesses if it is consistent with a co-moving (i.e., gravitationally bounded) object or a stationary background star with a null apparent on-sky motion (or with the motion of background objects as determined by e.g. Gaia astrometry). An example of both cases is presented in Appendix \ref{Appendix:cpm}. If the time between the two considered epochs is sufficiently large (typically, at least a year for the objects in the SHINE survey), this method can robustly separate the background objects from the bounded ones. By coupling SHINE and snapSHINE, we were able to identify 1180 physical (corresponding to 4217 detections) background sources and 26 physical bound sources (corresponding to 90 detections). 

\subsubsection{Direct Gaia detection}
\label{subsubsec:gaia}

As a second step, we used a tool to cross-match detected point sources in SPHERE images with the Gaia DR3 detected point sources catalog \citep{gaia_dr3}. The tool works as follows: it queries the Gaia DR3 archive around the target and retrieves all other Gaia detections in a given radius around the target (fixed at 8 arcsec for this analysis, slightly larger than the IRDIS field-of-view to avoid missing candidates). Then, the RA and DEC of all retrieved sources are propagated from the Gaia observations to the SPHERE epoch and converted in separation and parallactic angle to be easily identified in the S/N map. The tool also provides various information such as the proper motion, parallax, and fluxes of identified sources if available in the Gaia archive. An example is shown in Fig. \ref{fig:example_gaia_query}. This script allows the identification of background sources or binary systems even if only one observation is available. However, due to the intrinsic limitations of Gaia, we can not use this method for candidates fainter than a G magnitude of around 21 or too close to the star, typically less than 2 arcsec in projected separation \citep[see][for examples of non-applicable cases]{mu2_sco,Viswanath_hip81208,Chomez_Cb}. This tool allowed the identification of 34 additional background sources as well as 2 stellar binary, one of which was already identified in \cite{Bonavita_binaries}. 

\subsubsection{Color magnitude diagram}
\label{subsubsec:CMD}

For the remaining detected point sources, due to the already mentioned limitations of the follow-up observations, we have to use color-magnitude diagrams (CMDs) to classify them. This method is used for 1926\footnote{Note that because snapSHINE is conducted with the IRDIS BB\_H filter, the use of any CMD is impossible. Some sources at the edge of the FoV of snapSHINE are not visible in the SHINE epoch due to the large time interval between the two.} sources. The main goal of such a diagram is to try to disentangle potentially interesting sources from background contaminants to efficiently plan for follow-up observations. The design and usage of such a diagram have been described in detail in \cite{Bonnefoy_GJ504} and its initial usage by the GTO team in \cite{Langlois_shine}. As mentioned in \cite{Langlois_shine}, due to the IRDIS field-of-view, a large number of confirmed background sources (e.g.,  confirmed by using the common proper motion test) have been detected. Those background sources, mostly expected to be field M and K stars \citep{Parravano_2011}, fall within a particular area of the CMD. Thanks to their great number, we choose to re-define the exclusion area presented in \cite{Langlois_shine} taking advantage of the sheer number of background contaminants that we characterized in detail. This empirical distribution of bona-fide astrophysical contaminants is by construction representative of the true dispersion of contaminants in the CMD, accounting for the noise measurement but also for the natural color dispersion of background sources in SPHERE data.
The statistics of the color $C_{\text{color}}$ ($H2-H3$ or $K1-K2$) of confirmed background sources is computed (mean $\overline{C}_{\text{color}}$ and standard deviation $\sigma_{\text{color}}$) by bins of one magnitude to ensure enough data points. Figures \ref{fig:H23_stat_bckg}-\ref{fig:K12_stat_bckg} display the corresponding histograms with their Gaussian fit. 

\begin{figure}
    \centering
    \includegraphics[width = \linewidth]{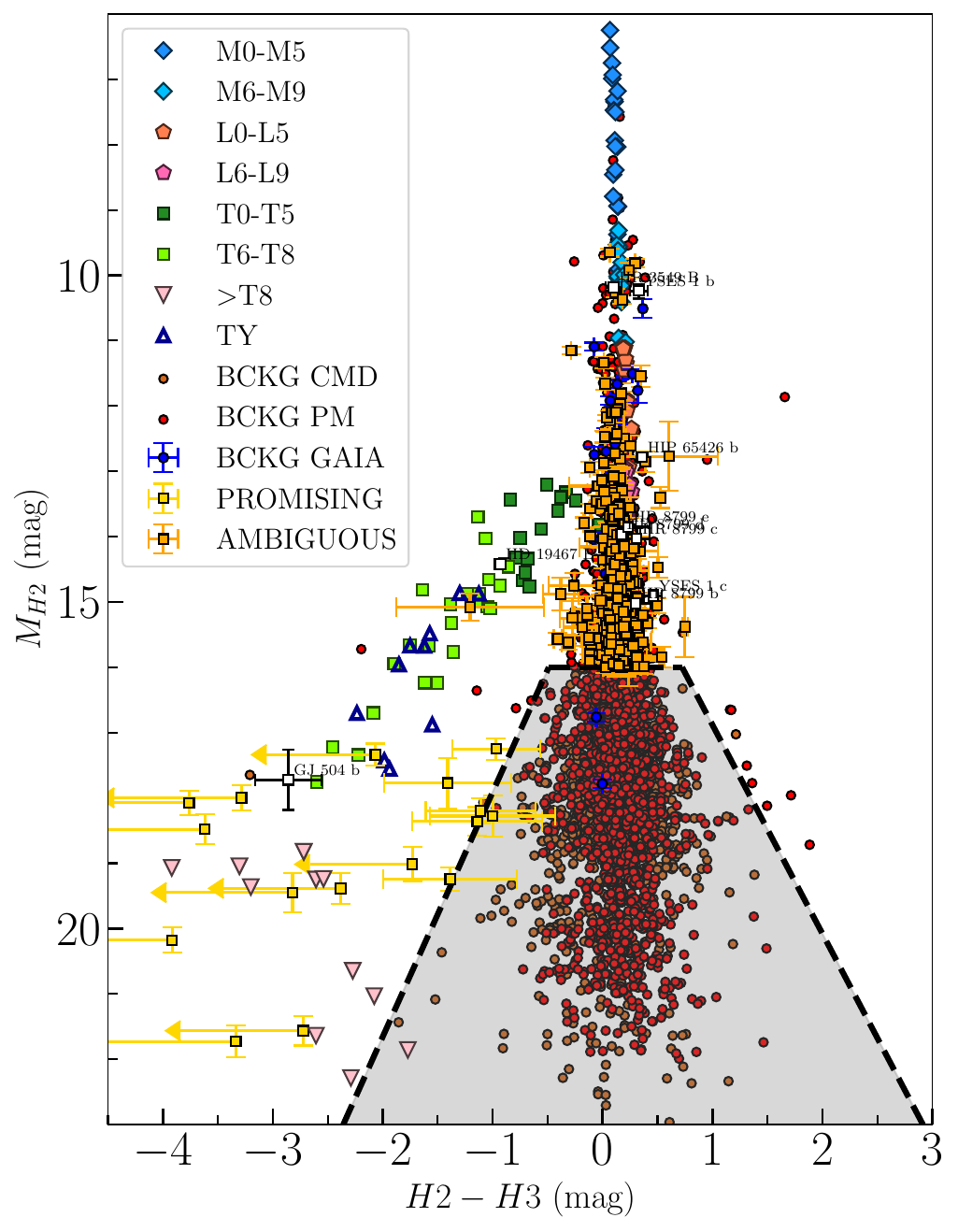}
    \caption{H2-H3 color-magnitude diagram with every non confirmed bound source detected during the SHINE survey. Error bars for background sources are removed for clarity purposes. The background exclusion area is indicated by the grey area and corresponds to a $5\sigma$ exclusion area.}
    \label{fig:CMD_H23}
\end{figure}

\begin{figure}
    \centering
    \includegraphics[width = \linewidth]{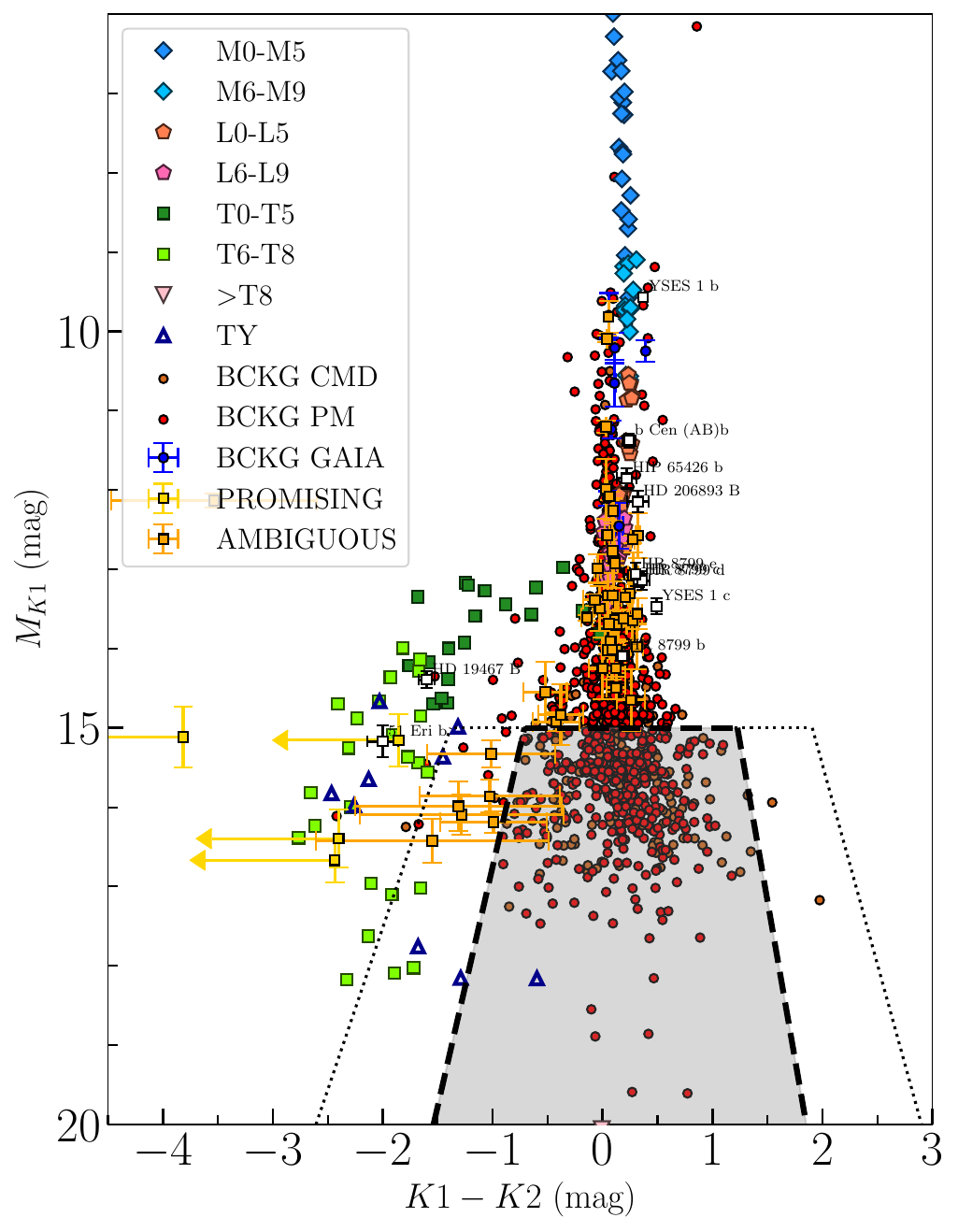}
    \caption{Same as Fig. \ref{fig:CMD_H23} but for the K12 IRDIS filter and a $3\sigma$ exclusion area. The 5$\sigma$ area is represented by the dotted lines. Sources between 3 and 5$\sigma$ are labeled as AMBIGUOUS (see explanation in text Sect. \ref{subsubsec:CMD}).}
    \label{fig:CMD_K12}
\end{figure}

The exclusion area is then defined by the area between \( \overline{C}_{\text{color}}\pm n \, \sigma_{\text{color}}\) for each bin of absolute magnitude. We choose here to set $n=5$. Those values are then linearly interpolated to obtain the final exclusion area. All sources whose position on the CMD falls in this exclusion area are then considered as background sources. This method is applied at magnitudes fainter than the L-T transition where the expected track of T and Y objects differs from M-L type objects. However, at magnitudes brighter than the L-T transition (seen at $M_{H2}=14$ in Fig. \ref{fig:CMD_H23}), background contaminants and sub-stellar companions overlap on the CMD. This is why we only use the background exclusion area below the L-T transition with an additional buffer of one magnitude below the absolute magnitude of HR 8799 b to avoid miss-classification of under luminous planets. The exclusion area thus defined can be seen in the CMDs in Figs. \ref{fig:CMD_H23}-\ref{fig:CMD_K12}. 

Additionally to the CMD, we associate, for every remaining detection where the exclusion area can be used, a Z-score, $Z$ aimed at identifying the most promising sources. We define this score $Z$ score by:
\begin{eqnarray}
    Z = \frac{|C_{\text{det}} - \overline{C}_{\text{bckg}}|}{\sigma_{\text{det}}} \,,
\end{eqnarray}
with $\sigma_{\text{det}}$ the error on the color of the source.

Although this technique is efficient with the H23 CMD, due to the increased thermal background noise in K1 and K2, the use of the K12 CMD is much more difficult. The error bars on each data point are typically larger than in H23, thus increasing the dispersion of the color $K1-K2$. As can be seen in Fig. \ref{fig:CMD_K12}, adopting the same method as in Fig. \ref{fig:CMD_H23} leads to a very wide background exclusion area, not suitable to disentangle between interesting sources and field contaminants. We choose then to mark the candidates between $3\sigma$ and $5\sigma$ as ambiguous because while within the expected dispersion range of background sources, they are also compatible with the CMD track of young T-type exoplanets. Additionally, all candidates sharing the same color and magnitude as M and L-type objects will also be marked as ambiguous as the CMD can not provide useful information to identify the most promising ones. We also want to stress that the detection indicated as promising based on colors should not be considered as new exoplanet detection but rather as the most promising candidates pending confirmation via second epoch observations.

\subsubsection{False positive identification}
\label{subsubsec:false_positives}

Given the unprecedented scale of this study and the number of observations, false positives are to be expected. Identifying false positives is a non-trivial task: it requires (at least) two observations and the confirming observation (i.e. the epoch where we are trying to re-detect a previously detected source) must be of higher quality to avoid any statistical noise fluctuation that would prevent the re-detection of a source.

Using this strategy, we were able to identify 191 false positives (100 on IRDIS, 91 on IFS). Their on-sky astrometry can be seen in Figs. \ref{fig:polar_plot_irdis} \& \ref{fig:polar_plot_ifs}, representing a polar plot (sep, PA) of all confirmed false positives for IRDIS and IFS, as well as their radial distribution. Their distributions present a slight increase at short separation in the speckle dominated area compared to a simple unbroken function of $r^2$. This excess is partly due to speckles causing extra false positives and partly because the quality of the noise estimation at these shorter separations is more affected when the FoV rotation is low, by lack of data diversity. This latter effect is a statistical counterpart of the self-subtraction that affects sources in ADI-based algorithms.

\cite{Chomez23_preparation} estimated the false positive rate for both IRDIS and IFS by using observations with an inverted PA rotation to destroy any physical signal, leaving only false positives. Using these results, we can estimate the expected number of false positives: 260 for IRDIS (0.4 per observation, $\sim$650 observations) and 162 for IFS (0.25 per observation, $\sim$650 observations). As demonstrated in \cite{Chomez23_preparation}, we can properly assess false positive rates under controlled conditions and the numbers then derived are representative of a standard SHINE observation. However, because of the variable observing conditions (see Fig. \ref{fig:fig_obs_cond}) and given the presence of a variety of astrophysical sources in our data, we acknowledge that the false positive rate will likely be higher, although this number is reduced by automated and human flagging of spurious detections. Given the large number of sources without a second epoch in the BCKG\_CMD and PROMISING categories, it is likely that some of them, especially at low S/N, are false positives. A more detailed discussion about false positives is available in Appendix \ref{Appendix:false_positive_polar_plots}.

\subsubsection{Summary of identifications}

\begin{figure}[t!]
    \centering
    \includegraphics[width = \linewidth]{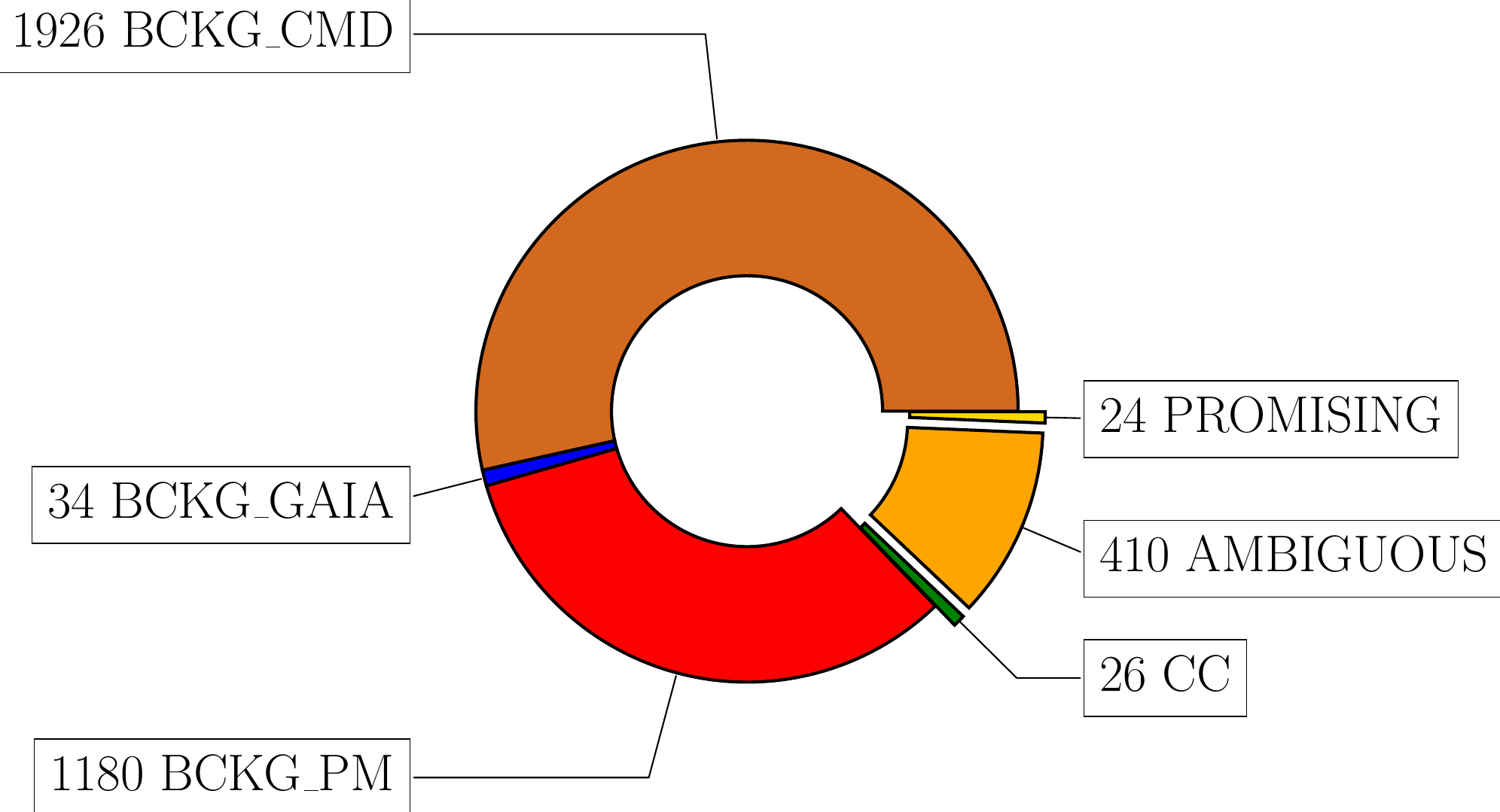}
    \caption{Pie chart summarizing the number of physical sources in each category for the SHINE F400 survey. CC stand for Confirmed Companion, BCKG\_GAIA for sources directly detected by GAIA and confirmed as background stars, BCKG\_PM for background stars confirmed via proper motion analysis and BCKG\_CMD, sources classified as background using a color magnitude diagram.}
    \label{fig:pie_chart}
\end{figure}

The final classification is as follows: "CC" for confirmed companion (stars, brown dwarfs, and planets) via proper motion, "BCKG\_PM" for background sources confirmed via proper motion, "BCKG\_CMD" for background sources identified via the CMD's exclusion area, "BCKG\_GAIA" for background sources identified with Gaia, "PROMISING" for promising candidates (e.g., candidates whose positions on the CMD is deemed interesting but no confirmation via proper motion can be made, either because no second epoch are available or the archival data are not of good enough quality) and finally "AMBIGUOUS" for ambiguous cases, where the CMD cannot distinguish between background and bound sources or the common proper motion test is non-conclusive or no reliable astro-photometry can be extracted. 
The pie chart in Fig. \ref{fig:pie_chart} summarizes the classification of the physical sources detected by SHINE according to the classes described above.

This analysis leaves 434 sources without a conclusive status (AMBIGUOUS + PROMISING). Most of those sources are likely background sources but their exact nature cannot be confirm at this stage, hence it is possible that several of these detections are real bound companions. Because the detection of any additional planet or brown dwarfs will have a strong impact on the derived statistics, mitigation strategies need to be implemented. While a detailed explanation of such strategies is out of the scope of this paper and will be fully explained in the paper VI, an example of such process can be found in \cite{Delorme_2024_beast} were the contrast map has been thresholded due to the presence of an unclassified source.

\begin{figure}[t!]
    \centering
    \includegraphics[width = \linewidth]{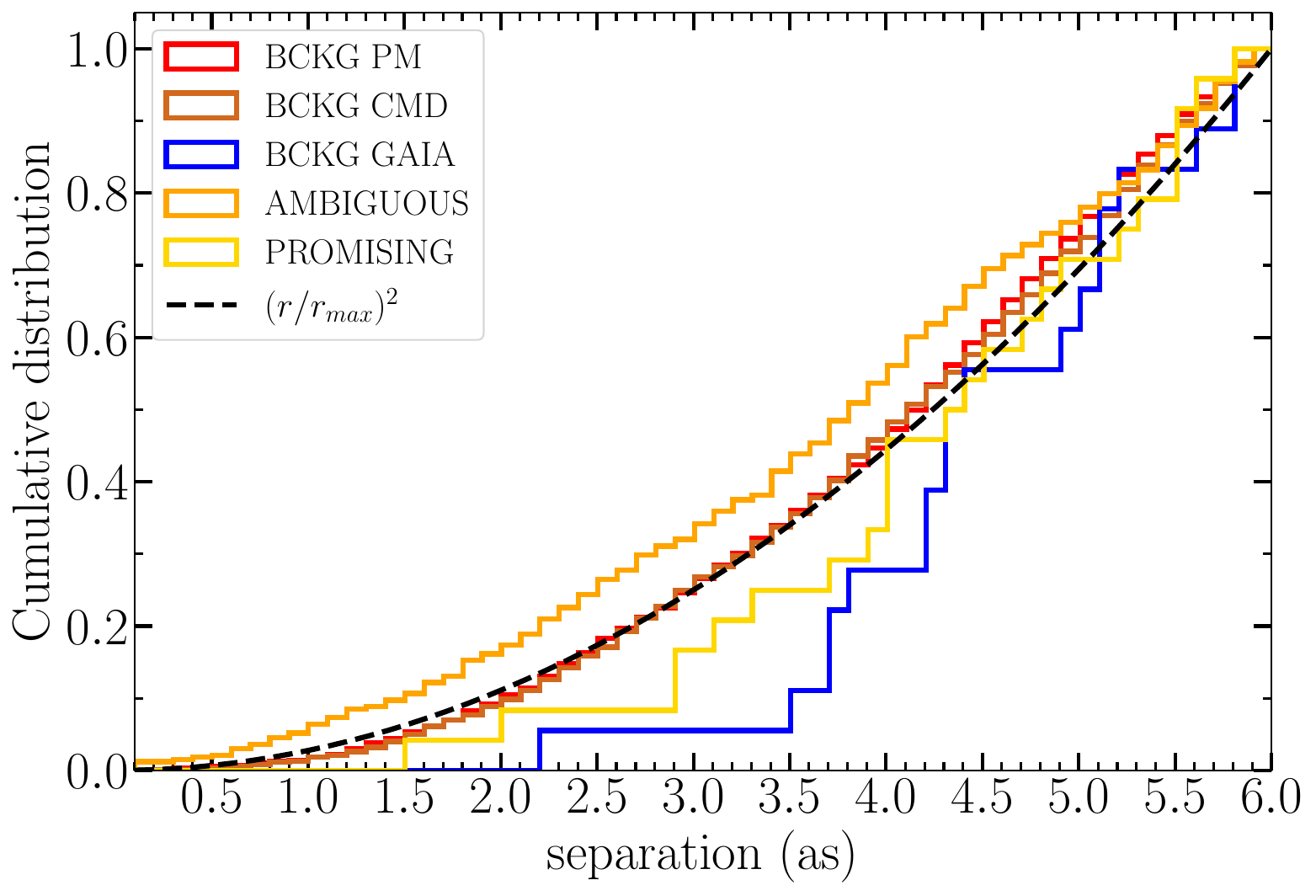}
    \caption{Cumulative normalized distribution of the separation of the sources according to their classes. Classes colors are identical to those in Figs. \ref{fig:CMD_H23}-\ref{fig:CMD_K12}-\ref{fig:pie_chart}.}
    \label{fig:cumulative_distrib}
\end{figure}

Figure \ref{fig:cumulative_distrib} show an histogram of the cumulative normalized distribution of the separation of all detected sources, divided according to their classification classes. Interestingly, the cumulative distribution of the sources classified as background with the proper motion (astrometric determination) match nearly perfectly the one of the background classified using the CMDs (photometric determination). To compare the two distributions we use the Kolmogorov-Smirnov test \citep[see e.g.][]{KS_test} with the null hypothesis of them being drawn from the same distribution. 
Using an $\alpha$ level of 0.05 (corresponding to 95\% confidence) we find that the two samples are drawn from the same distribution (p-value of 0.09). We can then be confident in our classification scheme using the CMD for those two categories. Additionally, they match the $(r/r_{max})^2$ which is the expected behavior: the number of sources visible in a field of view should increase quadratically with its radius.

As for the comparison between the AMBIGUOUS and BCKG\_PM distribution, the p-value is 0.006, suggesting that the two underlying distribution are different. We note that we are still probably suffering from two biases: (i) with the use of snapSHINE and due to the variable SHINE observing conditions, we are sometimes unable to confirm sources close to their stars and (ii) most of those ambiguous cases are around few stars. Given the limitation of SPHERE (both in term of sensitivity and inner working angle), it is unlikely that more than one or two sources per star are real bound companions (even this is an over-optimistic bound) and most of them are likely background stars. The number of individual stars harboring an ambiguous candidate is 69, which is 6 times lower than the number of ambiguous candidates. However, this excess (especially at short separation) can be seen as a hint about the presence of real physical bound companion in this class. We then carefully check the literature for existing archival observations taken outside the framework of SHINE (mainly with NACO, GPI and SPHERE) to vet ambiguous candidates. This excess is quite similar to the one shown in Fig. \ref{fig:polar_plot_irdis} and can be interpreted as an excess due to false positives. However, the S/N of the AMBIGUOUS sources, given their relatively high absolute magnitude, is typically larger than the one of the false positives. While is it likely that some false positives are hidden in this class, is it unlikely that they are the unique cause of this excess.

As for the two last classes (PROMISING and BCKG\_GAIA), the number of sources within is too small to compute any reliable statistics. We can still see the limitation in separation of the sources detectable by Gaia as more than 90\% of them are located at a separation greater than 3.5 arcsec.

\section{Conclusion}
\label{sec:ccl}

In this work, we analyzed in a homogeneous manner, over 700 datasets using advanced post-processing techniques. This survey, alongside GPIES \citep{Nielsen_GPIES_2019} and its recent reprocessing using \verb+PACO+ \citep{Squicciarini_GPI_2024}, is the largest survey ever conducted in the search for young giant planets in the solar neighborhood. We derived improved and statistically reliable contrast curves, critical to the upcoming study on demographics and formation mechanisms.

Combining SHINE and snapSHINE, over 7000 detections were reported, each needing a precise and thorough analysis to leave as few as possible sources in an ambiguous state. This tedious but essential analysis is mandatory to derive meaningful statistical constraints on the population and formation models. 

Such blind surveys, designed before the advent of Gaia-informed surveys, are still an essential component of planetary studies as they allow for the unbiased study of large statistical samples of stars. Although no new confirmed exoplanets have been found in this re-analysis and in the second half of the survey, due to the increased sensitivity, 24 additional promising sources have been unveiled. Those sources are either too far away from their host stars or, if bound, too light to cause any significant Gaia signal, needing further observations to confirm their nature.

Instrumental improvements such as SAXO+ \citep[an upgrade of the adaptive optic system of SPHERE;][]{Boccaletti_sphere_plus} and GPI 2.0 \citep{Chilcote_gpi_2} or improved observing techniques like dark hole on SPHERE \citep{Potier_dark_hole_2020, Potier_dark_hole_2022} will allow us to increase the sensitivity at small angular separations, unveiling potential new sub-stellar companions. 
Coupling such surveys as SHINE with other indirect techniques such as radial velocity and absolute astrometry (mainly with Gaia DR4, scheduled in 2026) will be precious to study the radial distribution of giant exoplanets from a fraction to hundreds of astronomical units.

\section*{Data Availability}
\label{sec:data_availability}

Tables with the astro-photometry of all detected sources by IRDIS and IFS and IFS spectrum of all sub-stellar companions are only available in electronic form at the CDS via anonymous ftp to cdsarc.u-strasbg.fr (130.79.128.5) or via \href{http://cdsweb.u-strasbg.fr/cgi-bin/qcat?J/A+A/}{http://cdsweb.u-strasbg.fr/cgi-bin/qcat?J/A+A/}.

\begin{acknowledgements}
This project has received funding from the European Research Council (ERC) under the European Union's Horizon 2020 research and innovation programme (COBREX; grant agreement n° 885593). 
SPHERE is an instrument designed and built by a consortium
consisting of IPAG (Grenoble, France), MPIA (Heidelberg, Germany),
LAM (Marseille, France), LESIA (Paris, France), Laboratoire Lagrange
(Nice, France), INAF - Osservatorio di Padova (Italy), Observatoire de
Genève (Switzerland), ETH Zürich (Switzerland), NOVA (Netherlands), ONERA
(France) and ASTRON (Netherlands) in collaboration with ESO. SPHERE
was funded by ESO, with additional contributions from CNRS (France),
MPIA (Germany), INAF (Italy), FINES (Switzerland) and NOVA (Netherlands).
SPHERE also received funding from the European Commission Sixth and Seventh
Framework Programmes as part of the Optical Infrared Coordination Network
for Astronomy (OPTICON) under grant number RII3-Ct-2004-001566 for
FP6 (2004-2008), grant number 226604 for FP7 (2009-2012) and grant number
312430 for FP7 (2013-2016). 
This work has made use of the High Contrast Data Centre, jointly operated by OSUG/IPAG (Grenoble),  PYTHEAS/LAM/CeSAM (Marseille), OCA/Lagrange (Nice), Observatoire de Paris/LESIA (Paris), and Observatoire de Lyon/CRAL, and is supported by a grant from Labex OSUG@2020 (Investissements d'avenir - ANR10 LABX56).
A.Z. acknowledges support from the ANID -- Millennium Science Initiative Program -- Center Code NCN2021\_080 and NCN2024\_001. 
T.B. acknowledges financial support from the FONDECYT postdoctorado project number 3230470.
We are grateful to the anonymous referee for the insightful comments provided during the peer-review, which largely contributed to raising the quality of the manuscript.

\end{acknowledgements}

\bibliographystyle{aa}
\bibliography{bibliography.bib}

\begin{appendix}

\section{Example of proper motion tests}
\label{Appendix:cpm}

\begin{figure*}
    \centering
    \includegraphics[width = \linewidth]{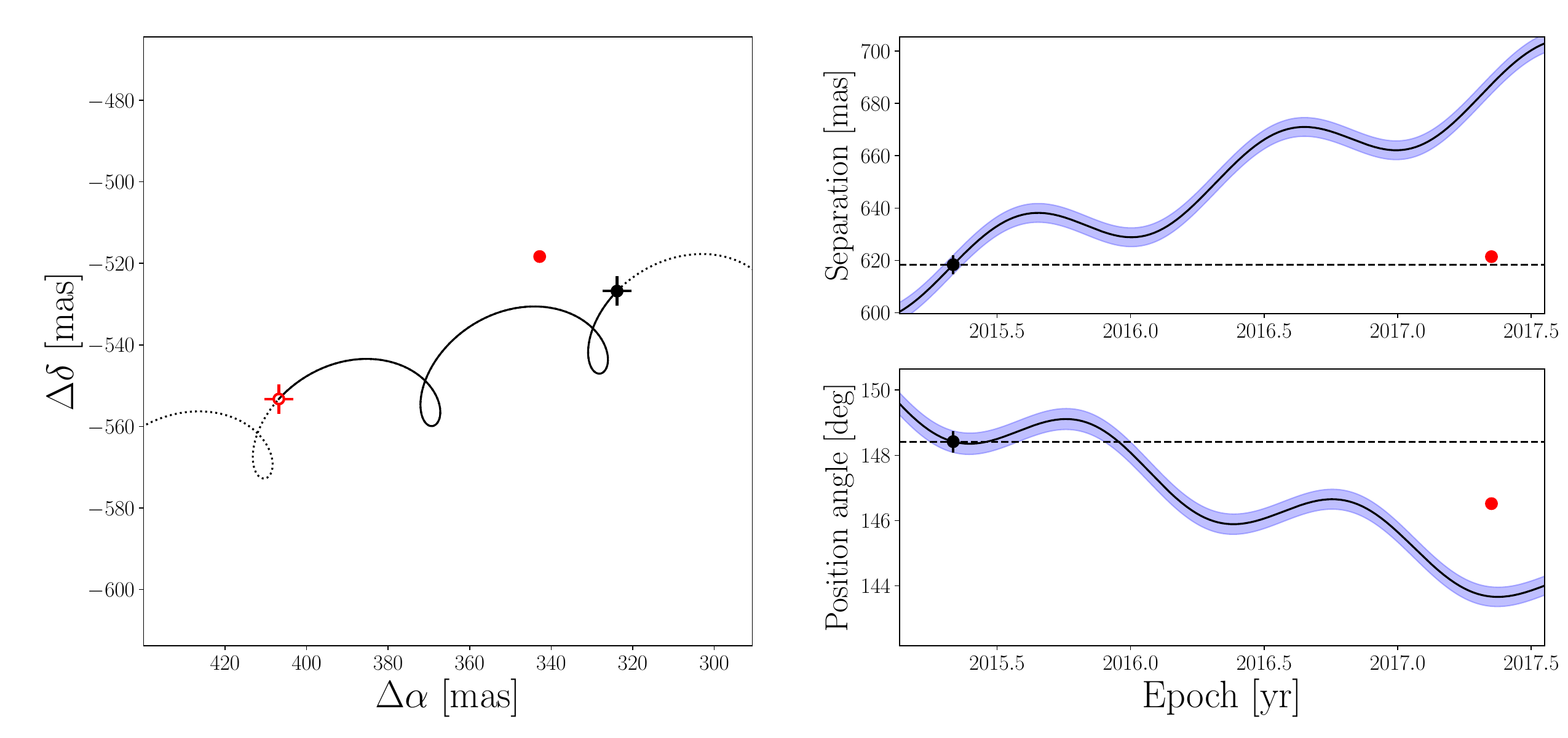}
    \caption{Common proper motion plot for the planet HD 95086 b. The position of the planet at the first epoch is represented by the black cross, the expected position at the second epoch, if background, by the red hollow cross, and the measured position at the second epoch by the plain red cross (Errorbars are included but due to the high S/N and the scale, they are not visible). The two red crosses are widely separated, displaying a motion non-compatible with a background source. Additionally, the motion of the planet is compatible with an orbital motion around the star.}
    \label{fig:HD95086b_cpm}
\end{figure*}

Example of two source identifications using the common proper motion test:

- HD 95086 b \citep[see e.g.][]{Chauvin_hd95086, Desgrange_HD95086} in Fig. \ref{fig:HD95086b_cpm} 

\begin{figure*}
    \centering
    \includegraphics[width = \linewidth]{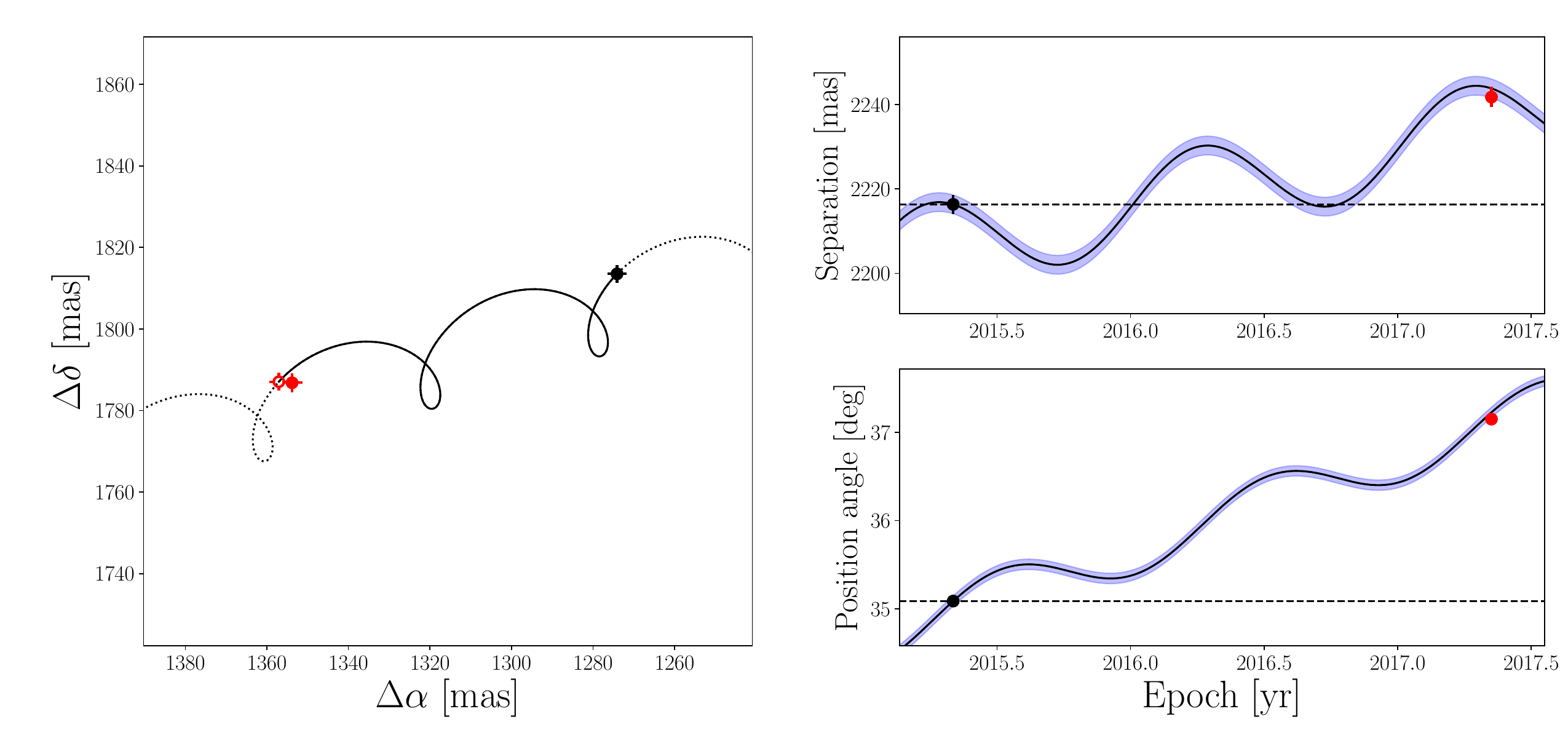}
    \caption{Same as Fig. \ref{fig:HD95086b_cpm} but for a background source presents in the IRDIS field-of-view. The two red crosses are compatible with each other, displaying a motion compatible with a background source.}
    \label{fig:HD95086bckg_cpm}
\end{figure*}

- a background source in Fig. \ref{fig:HD95086bckg_cpm}. 

This method uses the proper motion of the targeted star to disentangle co-moving sources (i.e., gravitationally bounded) from background sources, usually remote stars with little to no on-sky proper motion.

\newpage
\section{Example of a Gaia query to identify sources in the SPHERE field of view}
\label{Appendix:Gaia_cross_match}

Below is shown S/N map of an observation around the star TYC\_5164\_0567\_1 where three point-like sources were detected using PACO, labeled (1), (2), and (4). The Gaia query around this star (with a radius set at 8 arcsec to avoid missing sources) returns three detections indicated by the three green arrows. The properties of those detected stars are shown in Table \ref{tab:gaia_detected_sources}. The Gaia detection labeled (3) is outside the field-of-view of SPHERE, hence not visible in the image. The detection labeled (1) has a similar parallax and proper motion as the primary, hinting to a bound nature, confirmed by \cite{Bonavita_binaries}. Detection (2) is much further away than the primary, it is a background star. Finally, detection (4) is not seen by Gaia as its contrast is lower than the detection capabilities of Gaia, limited to sources with typically a G magnitude greater than 21.

\begin{table}[h!]
    \centering
    \caption{Detection parameters. (A) denotes the primary whose parameters have been retrieved via Simbad.}
    \begin{tabular}{c|c|c|c|c}
        \multirow{2}{*}{index} & pm\_ra  & pm\_dec & parallax & distance  \\
         & (mas/yr) & (mas/yr) & (mas) & (pc) \\
        \hline 
        (A) & 24.872 & -72.411 & 14.88 & 67.21 \\
        (1) & 22.902 & -79.035 & 14.95 & 66.89 \\
        (2) & -4.412 & -5.719 & 0.90 & 1107.08 \\
        (3) & 0.170 & -2.495 & 1.75 & 572.66 \\     
    \end{tabular}  
    \label{tab:gaia_detected_sources}  
\end{table}

\begin{figure}
    \centering 
    \includegraphics[width=\linewidth]{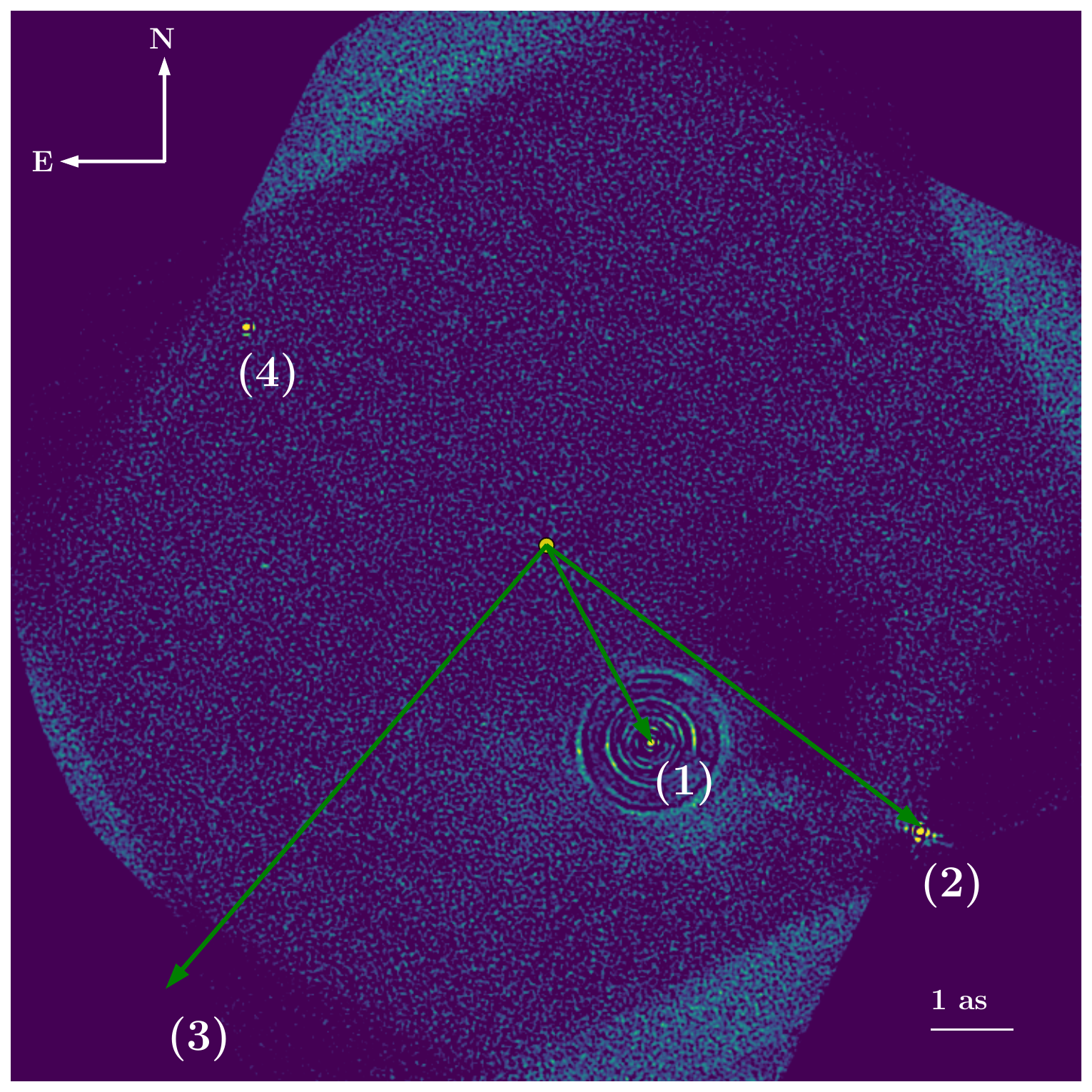}
    \caption{S/N map of the star TYC\_5164\_0567\_1 with three Gaia detected point-like sources labeled as (1), (2) and (4). The star is hidden behind the coronagraphic mask represented by the yellow circle at the center of the image. The three green arrows show direct Gaia detections.}
    \label{fig:example_gaia_query}
\end{figure}
\newpage
\section{Color distribution of confirmed background sources}
\label{Appendix:bckg_color_hist}

\begin{figure*}
    \centering 
    \includegraphics[width=\textwidth]{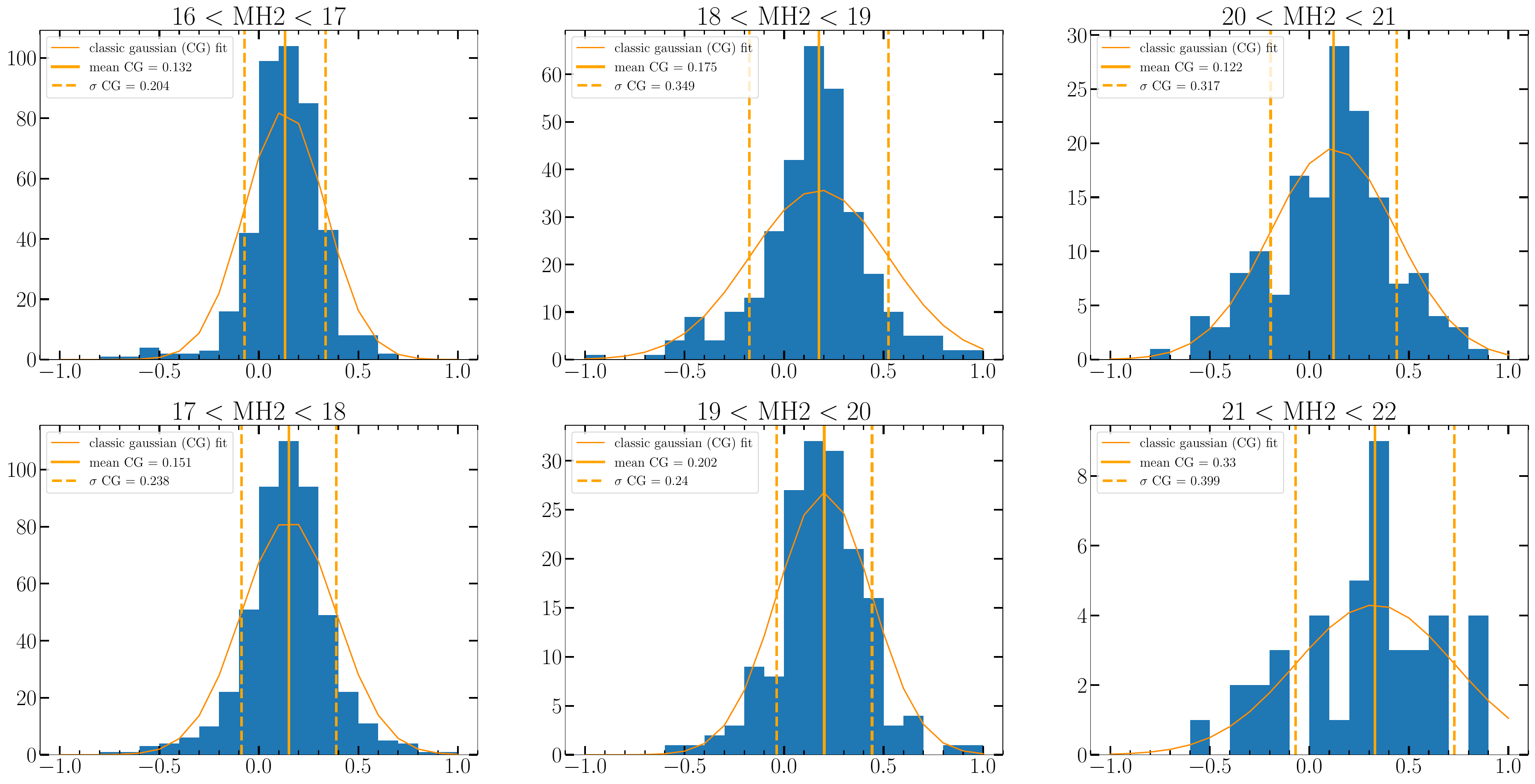}
    \caption{Histogram of the colors of confirmed background sources via proper motion by bin of magnitudes used to define the background exclusion area for the H23 CMD in Fig. \ref{fig:CMD_H23}.}
    \label{fig:H23_stat_bckg}
\end{figure*}

\begin{figure*}
    \centering 
    \includegraphics[width=\textwidth]{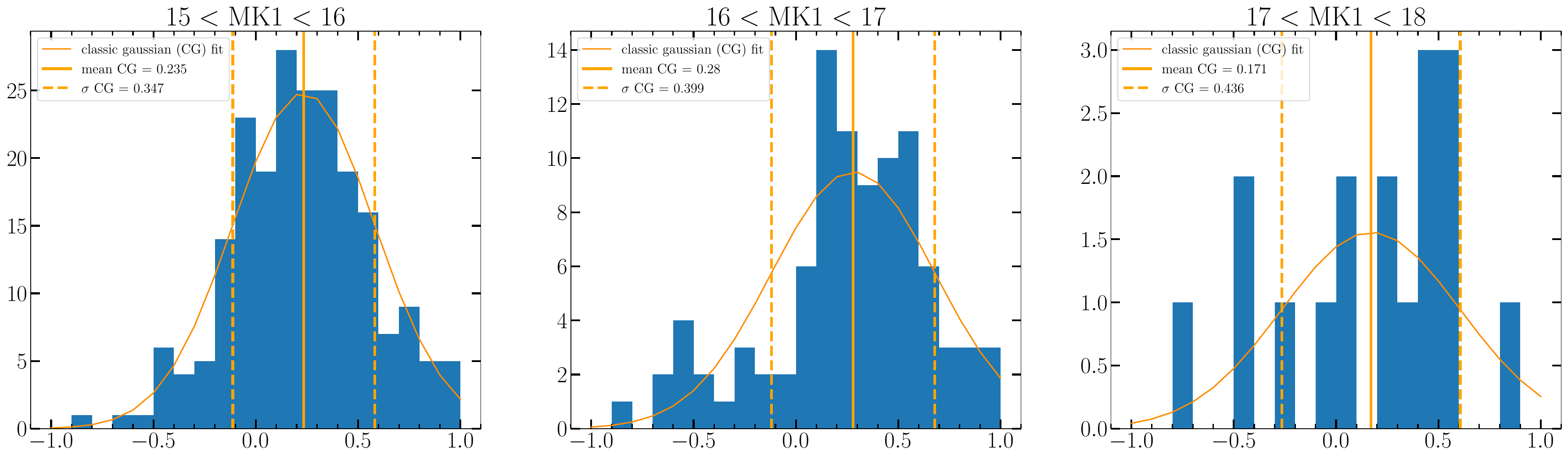}
    \caption{Same as \ref{fig:H23_stat_bckg} for the K12 CMD in Fig.\ref{fig:CMD_K12}.}
    \label{fig:K12_stat_bckg}
\end{figure*}

Histograms used to define the background exclusion area used in the CMDs in Fig. \ref{fig:CMD_H23} and \ref{fig:CMD_K12} are shown below for the \verb+IRDIFS+ configuration in Fig. \ref{fig:H23_stat_bckg} and for the \verb+IRDIFS-EXT+ in Fig. \ref{fig:K12_stat_bckg}.

\newpage
\section{False positives identification and distribution}
\label{Appendix:false_positive_polar_plots}

\begin{figure}[ht!]
    \centering 
    \includegraphics[width=\linewidth]{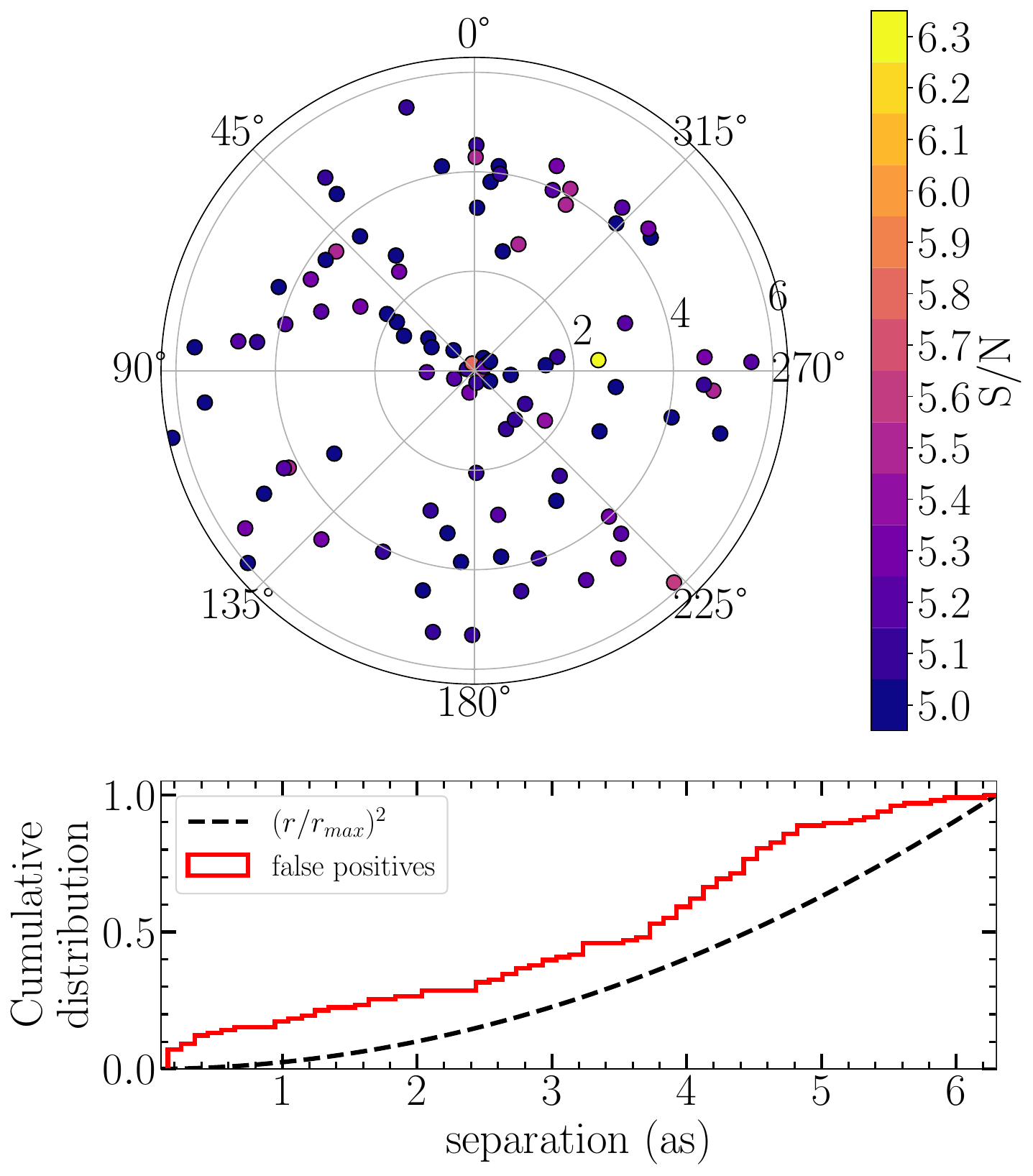}
    \caption{Polar plot showing the identified false positives on IRDIS and their radial distribution. The distribution is cut below 150 mas to ensure that the spectral leverage is enough to identify speckles.}
    \label{fig:polar_plot_irdis}
\end{figure}

\begin{figure}[ht!]
    \centering 
    \includegraphics[width=\linewidth]{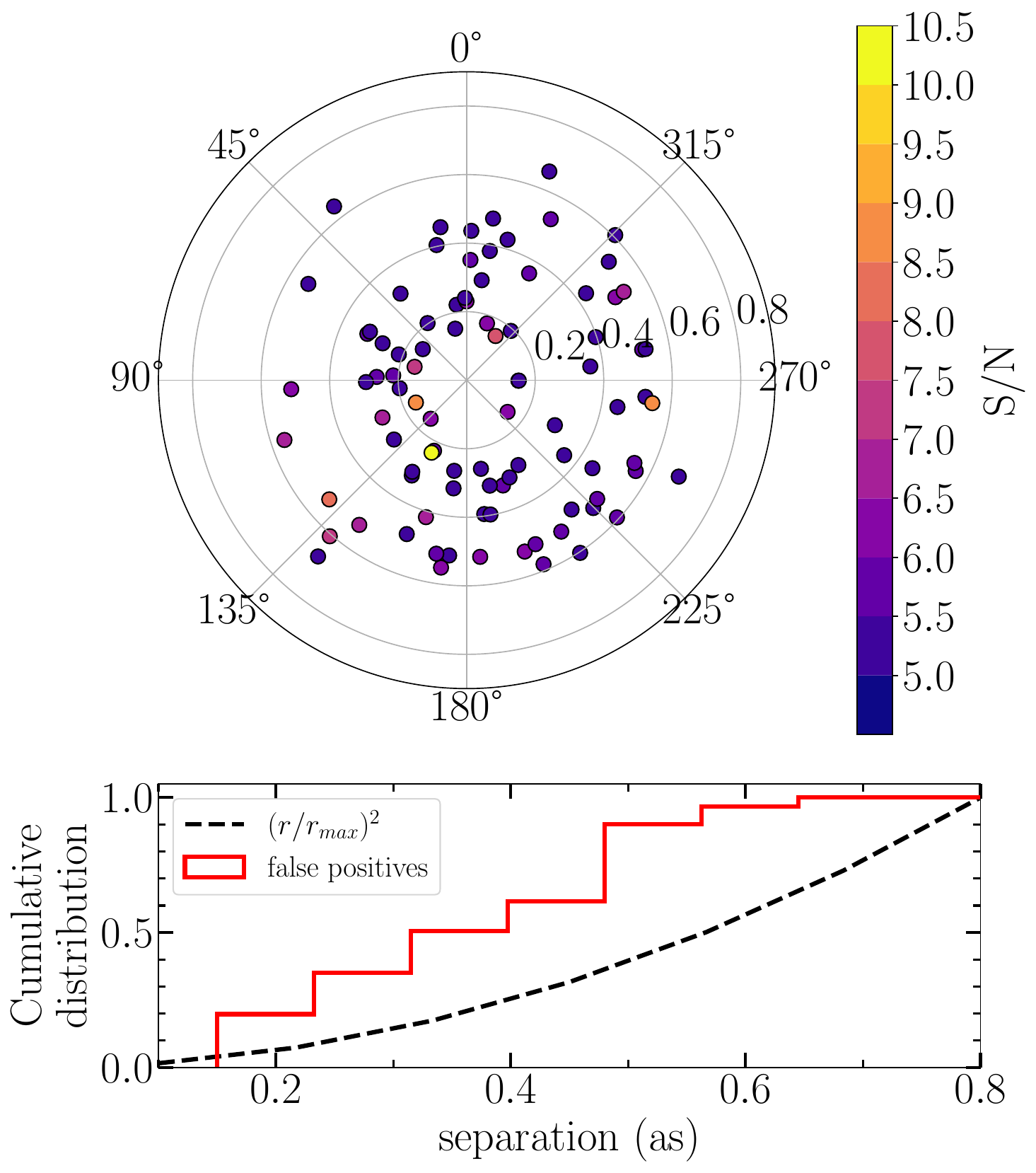}
    \caption{Polar plot showing the identified false positives on IFS and their radial distribution. The distribution is cut below 150 mas to ensure that the spectral leverage is enough to identify speckles.}
    \label{fig:polar_plot_ifs}
\end{figure}

The identification of false positives is an important subject for any massive analysis such as this one.
In summary, there are two complementary approaches to assess the statistical relevance of the claimed false alarm rate:
\begin{enumerate}
    \item Comparing the empirical occurrence of detected FWHM-wide sources, identified as false positives, to their expected theoretical occurrence rate at the prescribed detection threshold.
    \item Comparing the empirical distribution of the noise considered in the detection criterion, in the absence of astrophysical sources, to the expected theoretical distribution (i.e., a Gaussian with zero mean and unit variance).  In case of PACO the relevant value is the estimated SNR per pixel.
\end{enumerate}
Approach (1) is feasible when two epochs are available. It allows the clear identification of false positives, particularly if the second epoch is of higher quality, theoretically enabling the re-detection of sources with improved S/N. This eliminates random noise excursions that could lower the S/N and prevent the re-detection of a true astrophysical source. 
Where possible, we applied this method systematically on the survey, comparing the radial (angular) occurrence of false positives to the theoretical distribution (expected to follow a $r^2$ law if the noise in our SNR maps follows the same normal distribution in the speckle-noise dominated area and in the background-noise dominated area), see Sect. \ref{subsubsec:false_positives} and Figs. \ref{fig:polar_plot_irdis}-\ref{fig:polar_plot_ifs}. Our findings indicate a reasonable match between the two, especially given the heterogeneity of the survey, with a slight discrepancy at shorter angular separations within the speckle-dominated regime for IRDIS data. 
If we consider a pure thresholded analysis with PACO under 1" for IRDIS, we were able to identify 17 false positives in this area. This number is higher than the prediction in $r^2$ if the noise distribution is identical over the whole image. However, 8 of those 17 false positives are visible speckles identified thanks to the manual review of sources, leaving only 9 false positives, a number compatible with the expected one. The peculiar noise regime in the speckle-dominated area is therefore still responsible for a small increase in false positives, but this is partially compensated by reviewing detected sources.

Approach (2) is significantly more challenging, both to implement and then to interpret.
\begin{itemize}
    \item Considering jointly the distribution of the detection criterion on the whole survey by stacking all S/N maps introduces biases, even with detected sources masked,  from pathological cases (e.g., disks, smeared sources, or instrumental/reduction artifacts such as remanence leftovers). Such cases occur regularly across the hundreds of datasets we analyzed and result in false detections that are not solely representative of the noise distribution, as they arise from some form of signal rather than pure noise excursions. Even if not common, they can easily outnumber the small expected number of noise-induced excursions at four or five $\sigma$.
    \item Even stacking a carefully vetted subsample of good quality datasets will not directly yield the expected noise distribution because of the contribution to the histogram of astrophysical sources below the detection threshold and also above through stochastic amplification of a weak astrophysical source by random positive noise excursion. The latter is inherently unquantifiable, but the large number of background sources in our data above $5\sigma$, combined with an expected increase of such astrophysical contamination at fainter fluxes (probing a larger volume of the Milky Way), ensure that these sources will bias the derived distribution. 
\end{itemize}

As a result, the only reliable way to fully study the empirical false positive rate on the whole survey would be to reprocess all reductions using inverted parallactic angles. This would mitigates astrophysical signals by preventing constructive summation after derotation and stacking. However, systematically applying this method requires a large amount of computational resources and falls beyond the scope of this work given the tests already performed to ground the statistical relevance of our conclusions. Nonetheless, the following points should be emphasized:
\begin{itemize}
    \item Comparisons of the empirical distribution of the detection criterion (with reversed parallactic rotation) to the theoretical distribution have already been conducted in several case-by-case studies (under good to median observing conditions) to validate the statistical relevance of PACO, see e.g.  Figure 3 in \cite{Flasseur_paco}, Figures 2 and 4 in \cite{flasseur2020robustness}, and Figure 4 in \cite{Flasseur_asdi}. These studies consistently found a good match in all the cases analyzed.
    \item In addition, \cite{Chomez23_preparation} applied the same approach of inverted parallactic angles to estimate the false positive rate under good to median observing conditions. By extrapolating these results to the dataset analyzed in SHINE F400, we estimate 260 false positives for IRDIS and 162 for IFS. We acknowledge that this extrapolated value serves as a lower bound, as it accounts only for random noise excursions and not for the other sources of false positives discussed earlier. However, our final detections are not merely the result of applying a 5$\sigma$ threshold to the S/N maps. PACO incorporates several internal mechanisms to filter out pathological pixels and sources, and a manual review process during the analysis further eliminates additional spurious detections.
\end{itemize}

To summarize, we can reliably assess false positive rates under controlled observing conditions, and these estimates are representative of a standard SHINE observation. However, due to the variability in observing conditions and the presence of diverse astrophysical sources in our data, we acknowledge that the actual false positive rate on the whole survey is likely higher, and remains extremely challenging to evaluate more precisely. Nevertheless, this rate is mitigated by PACO's automated mechanisms and the additional manual flagging of spurious detections during the analysis.

\end{appendix}
\end{document}